\documentclass[a4paper,12pt,openany]{book}

\usepackage{indentfirst}
\usepackage[dvips]{graphicx}
\usepackage{amsmath,amssymb}
\usepackage[small]{caption}

\begin{document}

\thispagestyle{empty}
\begin{center}

{\Large 

Ph.D.~thesis

\vskip 1.5cm

{\bf Lattice QCD Anatomy\\
via the Energy-Momentum Component of Gluons}

\vskip 1.5cm

{\bf Arata Yamamoto}

\vskip 1.5cm

{\it Nuclear Theory Group\\
Department of Physics\\
Graduate School of Science\\
Kyoto University\\}
January 2011

}

\vskip 2cm

\begin{figure}[h]
\centering
\includegraphics[scale=0.8]{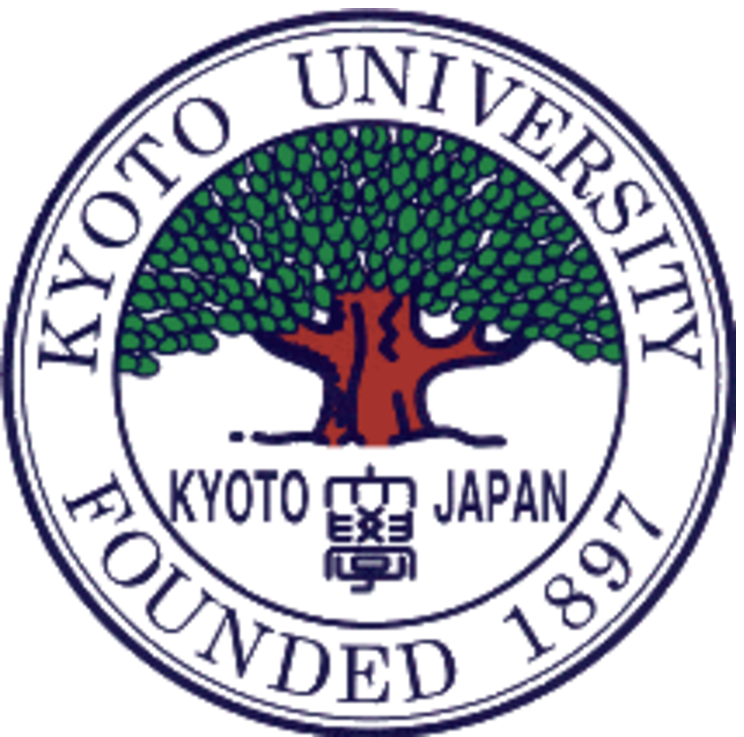}
\end{figure}

\end{center}

\chapter*{Abstract}
\thispagestyle{empty}
In this thesis, we perform the lattice QCD analysis via the energy-momentum component of gluons.
By introducing the momentum cutoff to the link variable, we investigate which energy-momentum components of gluons induce each QCD phenomenon.
We use the Landau gauge for the most part of the lattice QCD analysis.
In lattice QCD, we analyze color confinement, spontaneous chiral symmetry breaking, topological charge, and the related topics.
We also discuss several comparisons with effective theories.
As for color confinement, we calculate the quark-antiquark potential, the color flux tube, and meson masses.
From quantitative analysis, we find that color confinement is induced by the low-momentum component below 1.5 GeV.
As for spontaneous chiral symmetry breaking, we calculate the chiral condensate and the Dirac spectrum.
Spontaneous chiral symmetry breaking is induced by the broad low-momentum component which ranges even above 1.5 GeV.
The present result suggests that color confinement and spontaneous chiral symmetry breaking are induced by somehow different energy-momentum components of gluons.
As for topological charge, we calculate the topological charge density and the Dirac zero mode.
Topological charge is induced by the broad energy-momentum component, which is similar to spontaneous chiral symmetry breaking.

\clearpage

\tableofcontents
\thispagestyle{empty}

\chapter{Introduction}
\pagenumbering{arabic}
\thispagestyle{headings}

\subsubsection{Physics and scale}
In physics, the scale is ranged from the smallest size (the Planck scale) to the largest size (the universe).
There exist a huge number of physical phenomena.
These phenomena are described in various ways depending on their scales.
The small-scale phenomena are described by elementary particle physics, nuclear physics, and so on, and the large-scale phenomena are described by cosmology, astrophysics, and so on.
The scale is one of the most fundamental and important concepts in physics.

In quantum chromodynamics (QCD), there is a characteristic property describing the nontrivial appearance of a scale.
The QCD Lagrangian is composed of gluons and quarks as
\begin{eqnarray}
{\cal L}_{\rm QCD} = -\frac{1}{2}{\rm Tr} [ F^{\mu\nu}F_{\mu\nu} ] + \bar{q} [i\gamma_\mu D^\mu - m]q \ .
\end{eqnarray}
This Lagrangian has no dimensional parameter except for quark masses.
Thus, classical QCD is scale invariant in the chiral limit $m\to 0$.
After the quantization, however, scale invariance is violated by the trace anomaly, and a scale appears in a nontrivial manner.
This is called as dimensional transmutation.
In this way, many dimensional quantities are created in real QCD, such as ``mass gap'', vacuum condensates, and masses of hadrons.

\begin{figure}[t]
\begin{center}
\includegraphics[scale=0.5]{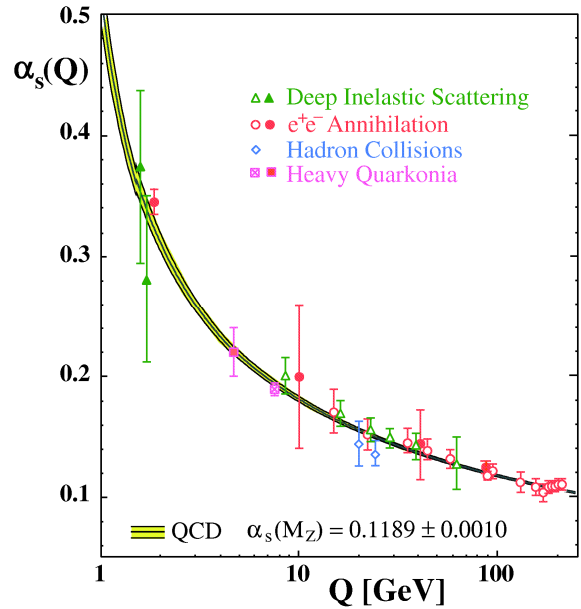}
\caption{\label{fig1}
QCD running coupling constant $\alpha_s(Q)$ \cite{Si07}.
}
\end{center}
\end{figure}

As a result of the dimensional transmutation, the QCD running coupling constant is a function of the energy scale (Fig.~\ref{fig1}).
At the one-loop perturbation theory, the QCD running coupling constant is
\begin{eqnarray}
\alpha_s(Q) = \frac{g^2(Q)}{4\pi} =\frac{1}{4\pi\beta_0 \ln (Q^2/\Lambda_{\rm QCD}^2)},
\end{eqnarray}
where $\beta_0=(11N_c-2N_f)/48\pi^2$ \cite{Gr73,Po73}.
For this reason, the behaviors of QCD phenomena are completely different among different energy scales.
At high energy or short distance, perturbative QCD is valid due to the asymptotic freedom.
At low energy or long distance, the interaction is strong coupling and nonperturbative effects are important.
The appearance of the energy scale enriches QCD phenomenology.

In this study, our question is ``which energy scale is relevant for {\it each} QCD phenomenon?''
It is often said that $\Lambda_{\rm QCD}$ is the typical scale of QCD.
However, our aim is not to determine $\Lambda_{\rm QCD}$.
$\Lambda_{\rm QCD}$ is not a quantitative scale of each phenomenon, but a typical scale of the whole theory.
Actually, even above $\Lambda_{\rm QCD}$, the running coupling constant is still large and the nonperturbative effect is important.
Since the energy scale is abstract concept, we have to define the energy scale in a more concrete way.
We define the {\it relevant energy scale} for a QCD phenomenon as the energy-momentum component of gluons inducing the phenomenon.
Thus, our question is rephrased as  ``which energy-momentum components of the gluon field induce each QCD phenomenon?''
By definition, such a relevant energy scale is different from $\Lambda_{\rm QCD}$.
It is nontrivial whether the relevant energy scales are the same or different between different phenomena, e.g., color confinement and spontaneous chiral symmetry breaking.
For a scale-dependent theory like QCD, the knowledge of the relevant energy scale would be useful for understanding physical phenomena.

To answer this question, we employ the numerical simulation of lattice QCD.
Lattice QCD is one of the most promising calculations for nonperturbative QCD.
In this thesis, we introduce a lattice QCD framework to determine which energy-momentum components of the gluon field induce QCD phenomena.
Using the introduced framework, we discuss the relevant energy scales, systematically, quantitatively, and nonperturbatively.

\subsubsection{Color confinement and spontaneous chiral symmetry breaking}
We focus on two of the most significant phenomena in QCD; color confinement and spontaneous chiral symmetry breaking.

In hadron phase, a color source, such as a quark or an antiquark, is confined in hadrons.
Color confinement symbolically reflects the strong interacting nature of QCD.
Phenomenologically, it has long been known that color confinement is well described by a ``string'' or ``flux tube'' \cite{Na74}.
The analytical derivation of color confinement has not yet been achieved because of the nonperturbative and non-Abelian nature of QCD.
It is not only an important theme in physics but also an unsolved problem in mathematics.
Lattice QCD studies have shown that color confinement is indeed realized, i.e., the Wilson loop obeys the area law \cite{Wi74,Cr7980}.
Color confinement is one of the main subjects in lattice QCD \cite{Ha99,Gr03}.

Although the QCD Lagrangian possesses chiral symmetry except for the quark mass term, this symmetry is spontaneously broken into its subgroup as
\begin{eqnarray}
{\rm SU}(N_f)_L \times {\rm SU}(N_f)_R \to {\rm SU}(N_f)_V.
\end{eqnarray}
Spontaneous chiral symmetry breaking is one dominant origin of mass in our world, and it is remarkably important in various aspects of hadron physics \cite{Na61,Hi91,Mi93,Ha94}.
Chiral symmetry itself is symmetry of quarks, not of gluons.
However, spontaneous breaking is dynamically induced by the nonperturbative interaction of gluons.
The gluon dynamics is essential for spontaneous chiral symmetry breaking.
The relation between the eigenmode of quarks and chiral symmetry breaking is known as the Banks-Casher relation \cite{Ba80}.
On the other hand, the relation between the energy-momentum component of gluons and chiral symmetry breaking is nontrivial.

The connection between color confinement and spontaneous chiral symmetry breaking is often discussed in QCD \cite{Ca79,Ma84,Su95,Mi95}.
At finite temperature, deconfinement and chiral symmetry restoring phase transitions occur at almost the same temperature \cite{Ko83,Po84,Fu86}.
We explore this connection in the viewpoint of the energy-momentum components of the gluon field at zero temperature.

\subsubsection{Outline of the thesis}
The outline of this thesis is as follows.
In Chapter 2, we briefly review basics of lattice QCD, and introduce our framework to analyze the relation between QCD phenomena and the energy-momentum component of the gluon field.
Using this framework, we determine the relevant energy scale in SU(3)$_c$ lattice QCD.
To investigate color confinement, we analyze the static quark-antiquark potential in Chapter 3, the color flux tube in Chapter 4, and the meson masses in Chapter 5.
To investigate spontaneous chiral symmetry breaking, we analyze the chiral condensate in Chapter 6, and the Dirac spectrum in Chapter 7.
Each simulation setup is shown in each chapter.
Finally, Chapter 8 is devoted to summary.

\chapter{Lattice QCD Formalism}
\thispagestyle{headings}
\section{Basics of lattice QCD}
\subsubsection{Partition function}
Lattice QCD is one of the most powerful tools to analyze nonperturbative QCD phenomena \cite{Cr81,Mo97,Ro92}.
In lattice QCD, the space-time is discretized as a hypercubic lattice with the lattice spacing $a$.
In the following, we consider the four dimensional space-time in the Euclidean metric.

The QCD partition function is given by
\begin{eqnarray}
Z_{\rm QCD} &=& \int DAD\bar{q}Dq\ e^{-S_{\rm gauge}-S_{\rm quark}}.
\end{eqnarray}
The lattice action is constructed so as to reproduce the continuum action in continuum limit $a\to 0$.
The expectation value of a physical quantity is computed by Monte Carlo simulation as
\begin{eqnarray}
\langle O \rangle &=& \frac{1}{Z_{\rm QCD}} \int DAD\bar{q}Dq \ O \ e^{-S_{\rm gauge}-S_{\rm quark}} \\
&\simeq& \frac{1}{N_{\rm conf}} \sum_{\rm conf} \ O.
\end{eqnarray}
The ensemble average is taken over the gauge configurations, which are generated with the weight function $\sim \exp (-S_{\rm gauge}-S_{\rm quark})$. 
The fermion action in the weight function is sometimes dropped for simplicity, which is called as the {\it quenched approximation}.
The quenched approximation corresponds to neglecting the quark loop effects.

\subsubsection{Gauge field}
In lattice QCD, the SU($N_c$) gauge field $A_\mu (x)$ is represented as the {\it link variable}
\begin{equation}
U_{\mu}(x)=e^{iagA_\mu (x)}.
\end{equation}
The minimum gauge invariant combination of the link variable,
\begin{eqnarray}
U_{\mu\nu}(x)=U_\mu(x)U_\nu(x+\hat{\mu})U_\mu^\dagger(x+\hat{\nu})U_\nu^\dagger(x),
\end{eqnarray}
is called as a {\it plaquette}.
The simplest lattice gauge action is written by this plaquette as
\begin{eqnarray}
S_{\rm gauge}&=&\sum _x s(x)\\
s(x)&=&\beta \sum_{\mu>\nu}\Bigl(1-\frac{1}{N_c}{\rm ReTr}U_{\mu\nu}(x)\Bigr) \label{eqAD}\\
\beta &=& \frac{2N_c}{g^2}.
\end{eqnarray}
This lattice gauge action corresponds to the continuum gauge action in continuum limit $a\to 0$.

The expectation value of a gauge variant operator is automatically zero without gauge fixing because of Elitzur's theorem \cite{El75}.
To calculate a gauge variant operator, gauge fixing is necessary.
Instead of the Faddeev-Popov method, gauge fixing is realized by imposing the gauge condition numerically.
For example, the Landau gauge is implemented by globally maximizing the quantity
\begin{eqnarray}
F_L[U] \equiv \sum_{x}\sum_{\mu} {\rm ReTr} U_\mu (x),
\end{eqnarray}
by the gauge transformation.
In continuum limit, this condition is equivalent to minimizing 
\begin{eqnarray}
\int d^4x {\rm Tr}\{ A_\mu (x)^2 \},
\end{eqnarray}
and it is a sufficient condition for the local condition $\partial_\mu A_\mu (x)=0$.

\subsubsection{Fermion field}
The fermion action is the bilinear form of the quark field,
\begin{eqnarray}
S_{\rm quark}=\sum_{x,y} \bar{q}(x) D q(y).
\end{eqnarray}
If one naively discretizes the continuum Dirac operator $D = \gamma_\mu \partial_\mu - ig\gamma_\mu A_\mu$, the lattice Dirac operator suffers from the {\it doubling problem} \cite{Ni81}.
Several ways have been proposed to avoid the doubling problem.

One familiar lattice fermion is the Wilson fermion \cite{Wi7577}.
The Wilson Dirac operator is
\begin{eqnarray}
D&=&\delta _{x,y} -\kappa \sum_\mu \bigl \{ (1-\gamma_\mu )U_\mu(x)\delta_{x+\hat{\mu},y}+(1+\gamma_\mu )U_\mu^\dagger(x-\hat{\mu})\delta_{x-\hat{\mu},y}\bigr \}, \label{eqDwi}\\
\kappa&=&\frac{1}{2ma+8}.
\end{eqnarray}
The Wilson Dirac operator includes the $O(a^2)$ term which breaks chiral symmetry explicitly.
The degree of freedom of flavor can be introduced without restriction.

Another familiar lattice fermion is the staggered fermion \cite{Ko75,Su77}.
The staggered Dirac operator in the spinorless basis is
\begin{eqnarray}
D &=&m\delta _{x,y} + \frac{1}{2} \sum_\mu \eta_\mu (x) [U_\mu(x) \delta_{x+\hat{\mu},y} - U^\dagger_\mu(x-\hat{\mu})  \delta_{x-\hat{\mu},y} ]\ ,\label{eqDst}\\
\eta_\mu (x) &=& (-1)^{x_1+ \cdots + x_{\mu-1}}\ .
\end{eqnarray}
The staggered Dirac operator describes the degenerate $N_f=4$ Dirac fermion in continuum limit.
At finite lattice spacing, however, it includes an $O(a^2)$ term which violates the flavor symmetry, and the symmetry is reduced to ${\rm U}(1)_V \times {\rm U}(1)_A$.
Compared to the Wilson fermion, the staggered fermion has the ${\rm U}(1)_A$ subgroup of the chiral symmetry which is exact on the lattice, but is restricted in the number of flavor.

The above two fermions break the full chiral symmetry.
As a result, the anti-commutation relation $\{ D, \gamma_5 \} = 0$, which holds in continuum QCD, is not satisfied.
There are other lattice fermions with better chiral property, such as the overlap fermion \cite{Ne98}, the domain-wall fermion \cite{Ka92}, and so on.

\section{Momentum cutoff}
To investigate the relevant energy scale, we introduce a lattice framework to remove some region of momentum space by a momentum cutoff.
By observing how a physical quantity is affected by the momentum cutoff, we can determine the role of the removed energy-momentum component.
This concept has been applied, for example, in the Swinger-Dyson approach \cite{Ii05}.
We introduce this concept to lattice QCD, and analyze the energy-momentum component of the gauge field.
The framework is applicable to both quenched and full QCD in the same way.

The framework is formulated as the following five steps.

\subsubsection{Step 1}
The SU(3)$_c$ link variable $U_{\mu}(x)$ is generated by Monte Carlo simulation.
Boundary conditions for the link variable are taken to be periodic.
As explained below, the link variable must be fixed with a certain gauge.
In the following, we mainly use the Landau gauge for the numerical calculation, unless otherwise stated.

\subsubsection{Step 2}
The momentum-space link variable ${\tilde U}_{\mu}(p)$ is obtained by the Fourier transformation, as
\begin{eqnarray}
{\tilde U}_{\mu}(p)=\frac{1}{V}\sum_{x} U_{\mu}(x)\exp(i \sum_{\nu} p_\nu x_\nu ),
\end{eqnarray}
where $V$ is the lattice volume.
The momentum-space lattice has the same structure as the coordinate-space lattice.
Its lattice volume is $V$, and the boundary conditions are periodic.
The momentum-space lattice spacing is given by
\begin{eqnarray}
a_p = \frac{2\pi}{La},
\end{eqnarray}
where $L$ is the number of lattice sites in each direction.
The momentum-space lattice spacing corresponds to the minimum unit of momentum, and it has mass dimension.
The first Brillouin zone of the momentum space is $(-\pi/a, \pi/a ]$.

\begin{figure}[t]
\begin{center}
\includegraphics[scale=0.5]{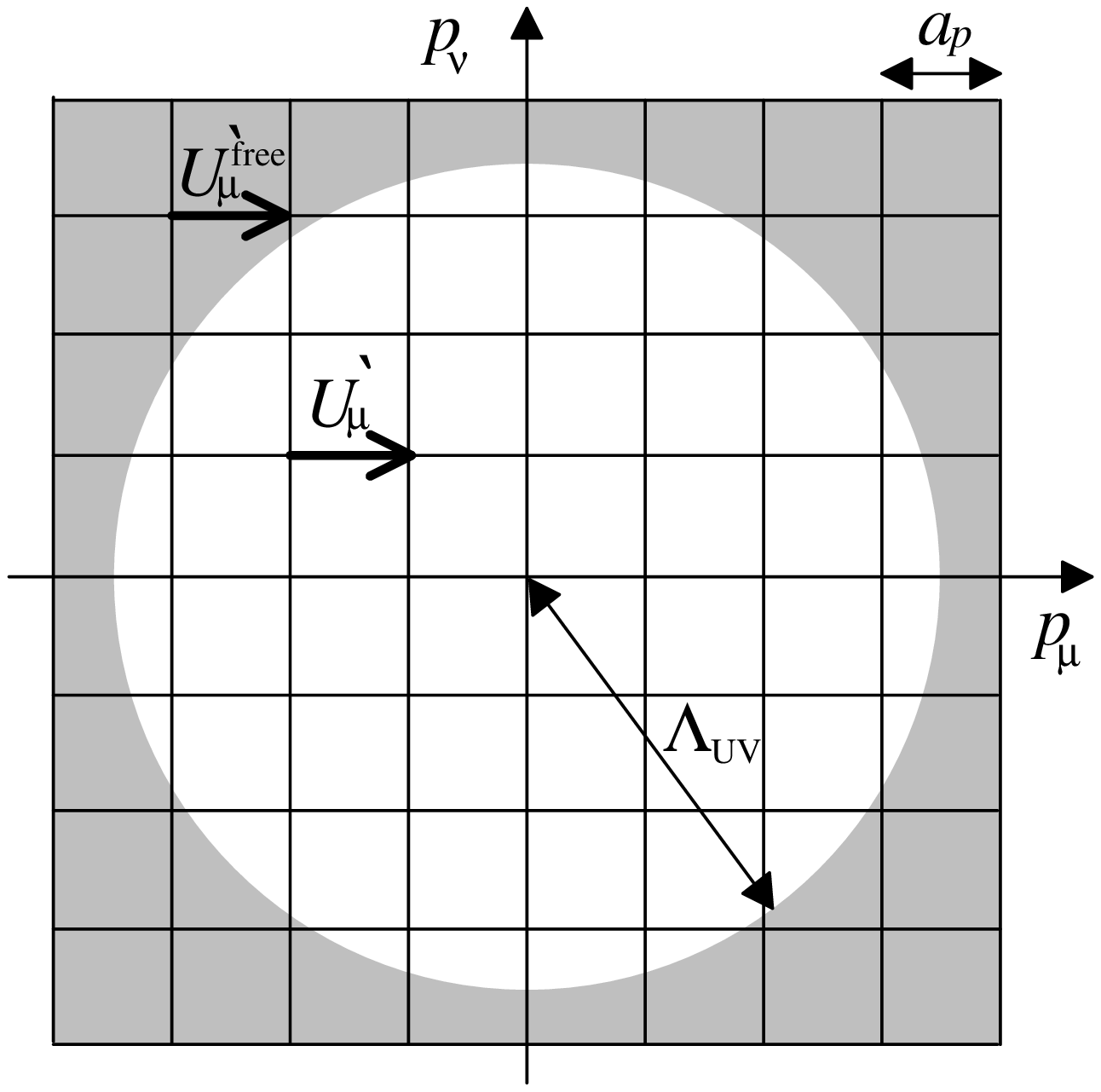}
\includegraphics[scale=0.5]{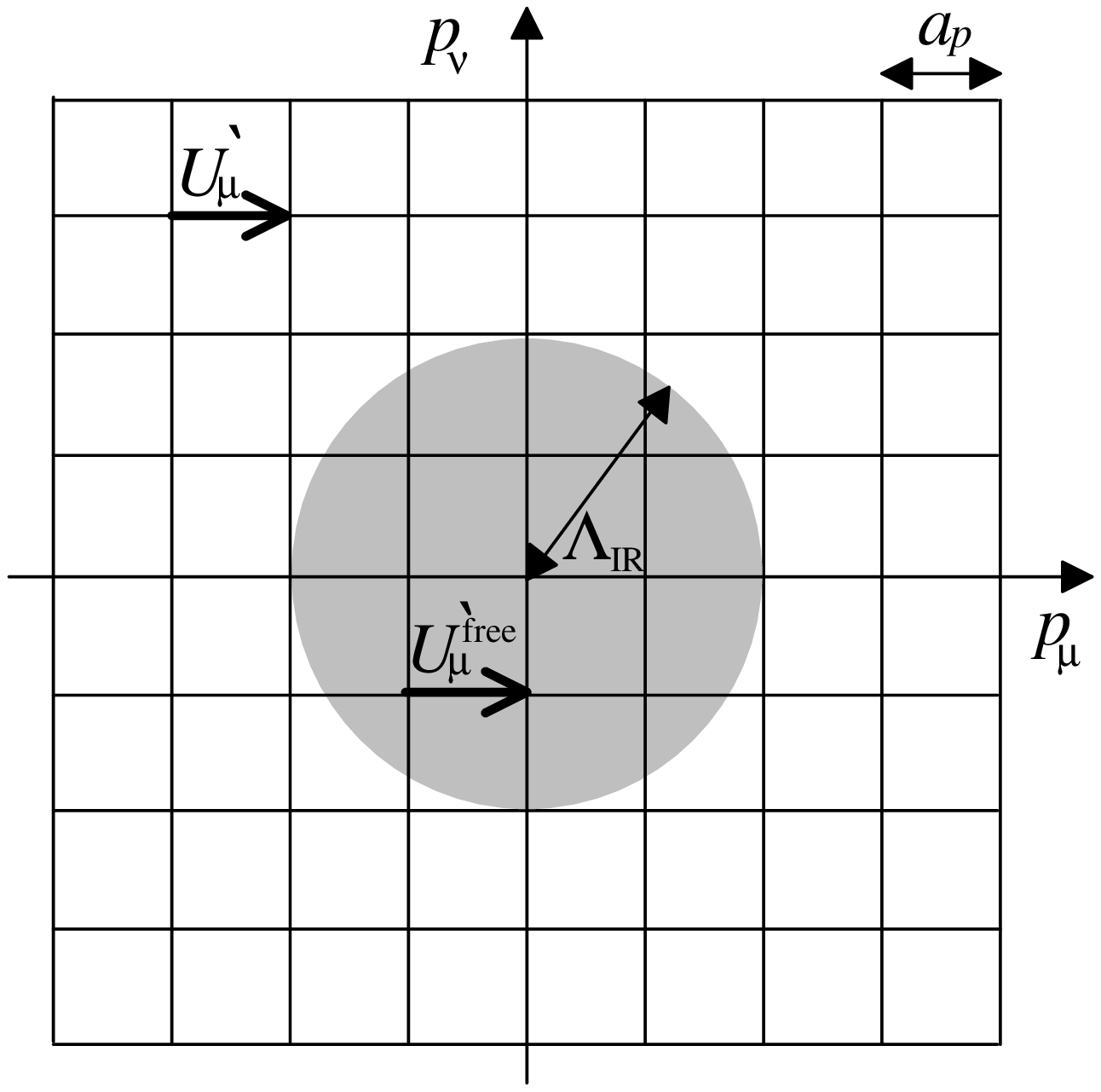}
\caption{\label{fig2-1}
The schematic figures of momentum space.
The shaded regions are the cut regions by the UV cutoff $\Lambda_{\rm UV}$ (left) and the IR cutoff $\Lambda_{\rm IR}$ (right).
The momentum-space lattice spacing is $a_p= 2\pi/La$.
}
\end{center}
\end{figure}

\subsubsection{Step 3}
Some components of ${\tilde U}_{\mu}(p)$  are removed by introducing a momentum cutoff.
In the cut region, the momentum-space link variable is replaced by the free-field link variable,
\begin{equation}
\label{eq2}
{\tilde U}_{\mu}^{\Lambda}(p)= \Bigg\{
\begin{array}{cc}
{\tilde U}^{\rm free}_{\mu}(p) & ({\rm cut \ region})\\
{\tilde U}_{\mu}(p) & ({\rm other \ region}),
\end{array}
\end{equation}
where
\begin{equation}
{\tilde U}^{\rm free}_{\mu}(p)=\frac{1}{V}\sum_x 1 \exp(i {\textstyle \sum_\nu} p_\nu x_\nu)=\delta_{p0}.
\end{equation}

The concrete form of the momentum cutoff can be taken arbitrarily.
For example, one natural choice is the cutoff by the four-momentum length $\sqrt{p^2}=\sqrt{\sum_\mu p_\mu p_\mu}$, which corresponds to a simple momentum cutoff in continuum theory.
The ultraviolet (UV) cutoff $\Lambda_{\rm UV}$ is introduced as 
\begin{equation}
{\tilde U}_{\mu}^{\Lambda}(p)= \Bigg\{
\begin{array}{cc}
{\tilde U}_{\mu}(p) & (\sqrt{p^2} \le \Lambda_{\rm UV})\\
0 & (\sqrt{p^2} > \Lambda_{\rm UV}),
\end{array}
\end{equation}
and the infrared (IR) cutoff $\Lambda_{\rm IR}$ is introduced as
\begin{equation}
{\tilde U}_{\mu}^{\Lambda}(p)= \Bigg\{
\begin{array}{cc}
\delta_{p0} & (\sqrt{p^2} < \Lambda_{\rm IR})\\
{\tilde U}_{\mu}(p) & (\sqrt{p^2} \ge \Lambda_{\rm IR}).
\end{array}
\end{equation}
The schematic figures are shown in Fig.~\ref{fig2-1}.

\subsubsection{Step 4}
The coordinate-space link variable with the momentum cutoff is obtained by the inverse Fourier transformation as
\begin{eqnarray}
U'_{\mu}(x)=\sum_{p} {\tilde U}_{\mu}^{\Lambda}(p)\exp (-i \sum_{\nu} p_\nu x_\nu ).
\end{eqnarray}
Since $U'_{\mu}(x)$ is not an SU(3)$_c$ matrix in general, $U'_{\mu}(x)$ must be projected onto an SU(3)$_c$ element $U^{\Lambda}_{\mu}(x)$.
The projection is realized by maximizing the quantity
\begin{eqnarray}
{\rm ReTr}[\{ U^{\Lambda}_{\mu}(x) \}^{\dagger}U'_{\mu}(x)].
\end{eqnarray}
By this projection, we obtain the coordinate-space link variable $U^{\Lambda}_{\mu}(x)$ with the momentum cutoff, which is an SU(3) matrix and has the maximal overlap to $U'_{\mu}(x)$.

\subsubsection{Step 5}
The expectation value of the operator $O$ is calculated from the link variable $U^{\Lambda}_{\mu}(x)$ instead of $U_{\mu}(x)$, i.e., $\langle O[U^\Lambda]\rangle$ instead of $\langle O[U]\rangle$.

\subsubsection{ }
We repeat these five steps with different values of the momentum cutoff, and observe the dependence of a physical quantity on the momentum cutoff.
Then, we can determine which energy-momentum component of the gluon field is relevant for the physical quantity.
Technically, we only have to replace $U^{\Lambda}_{\mu}(x)$ instead of $U_{\mu}(x)$, and its operation (mainly the fast Fourier transformation) is numerically easy compared to other parts of the simulations.
This framework is applicable to almost all lattice QCD calculations, and widely useful for QCD phenomena.

We comment on several points about this framework in the following.
\begin{enumerate}
\item 
In general, since the gauge transformation is nonlocal in momentum space, the energy-momentum component of the gauge field is a gauge-dependent concept.
The gauge fixing is needed and the result depends on the gauge choice.
If we do not fix the gauge, the resulting expectation value is zero, even in the case of a gauge-invariant operator.

\item 
It is possible that the projection in Step 4 contaminates the original condition on the momentum cutoff.
To evaluate how the projection affects link variables, we calculated $U^{\Lambda\Lambda}_{\mu}(x)$ by repeating Steps 2-4 once again to $U^{\Lambda}_{\mu}(x)$, and check the overlap between them, $\frac{1}{3}{\rm ReTr}[\{ U^{\Lambda}_{\mu}(x) \}^{\dagger}U^{\Lambda\Lambda}_{\mu}(x) ]$. 
The overlap is found to be almost unity.
For example, the average deviation from unity is about 0.1\% at $\Lambda_{\rm IR}=1.5$ GeV.
Thus, we can expect that the projection does not significantly affect the original momentum cutoff.

\item
In general gauges, since the gauge fixing and the momentum cutoff do not necessarily commute, it is nontrivial whether these two operations are satisfied simultaneously.
In other words, $U^{\Lambda}_{\mu}(x)$ can deviate from the gauge fixing condition which is originally imposed on $U_{\mu}(x)$.
We have numerically checked that, in the case of the Landau gauge, $U^{\Lambda}_{\mu}(x)$ almost completely  satisfies the Landau gauge fixing condition, i.e., $F_L[U^{\Lambda}]$ is maximized.

\item The Landau gauge is known to suffer from the Gribov copy problem \cite{Gr78}.
We numerically estimated the systematic uncertainty from the Gribov ambiguity by random gauge transformation.
In the cases of the Wilson loop and plaquette, its systematic error is smaller than the statistical error.
\end{enumerate}

\chapter{Interquark Potential}
\thispagestyle{headings}

The interquark potential is one of the most fundamental quantities in QCD and hadron physics.
We here calculate the static quark-antiquark potential.
The quark-antiquark potential, so-called the Cornell potential \cite{Ei78}, is given as
\begin{eqnarray}
V(R)=\sigma R -\frac{A}{R} +C,
\label{VQQ}
\end{eqnarray}
where $R$ is the interquark distance.
The physical value of the string tension $\sigma$ is approximately 0.89 GeV/fm, and the Coulomb coefficient $A$ is approximately 0.26.
The constant $C$ depends on the regularization.
The quark-antiquark potential includes both perturbative and nonperturbative properties.
In short range, it is dominated by the perturbative one-gluon-exchange Coulomb potential.
In long range, it is dominated by the nonperturbative linear confinement potential.

The quark-antiquark potential is extracted from the expectation value of the Wilson loop, which is a gauge-invariant path-ordered product of link variables along a loop.
For statistical improvement, we used the APE smearing method \cite{Al87}.
The simulation parameters are listed in Table \ref{tab3-1}.

\begin{table}[h]
\renewcommand{\tabcolsep}{1pc} 
\renewcommand{\arraystretch}{1} 
\caption{\label{tab3-1}
Simulation parameters of the gauge configurations.
The lattice coupling $\beta=2N_c/g^2$, the lattice volume $V$, the coordinate-space lattice spacing $a$, the momentum-space lattice spacing $a_p$, and the configuration number $N_{\rm conf}$ are listed.
}
\begin{center}
\begin{tabular}{cccccc}
\hline\hline
& $\beta$ & $V$ [$a^4$] & $a$ [fm] & $a_p$ [GeV] & $N_{\rm conf}$\\
\hline
Quenched & 5.7 & $16^4$ & 0.19 & 0.41 & 50\\
Quenched & 5.8 & $16^4$ & 0.14 & 0.55 & 50\\
Quenched & 6.0 & $16^4$ & 0.10 & 0.77 & 50\\
\hline\hline
\end{tabular}
\end{center}
\end{table}

\section{Interquark potential with the momentum cutoff}

\begin{figure}[p]
\begin{center}
\includegraphics[scale=1.5]{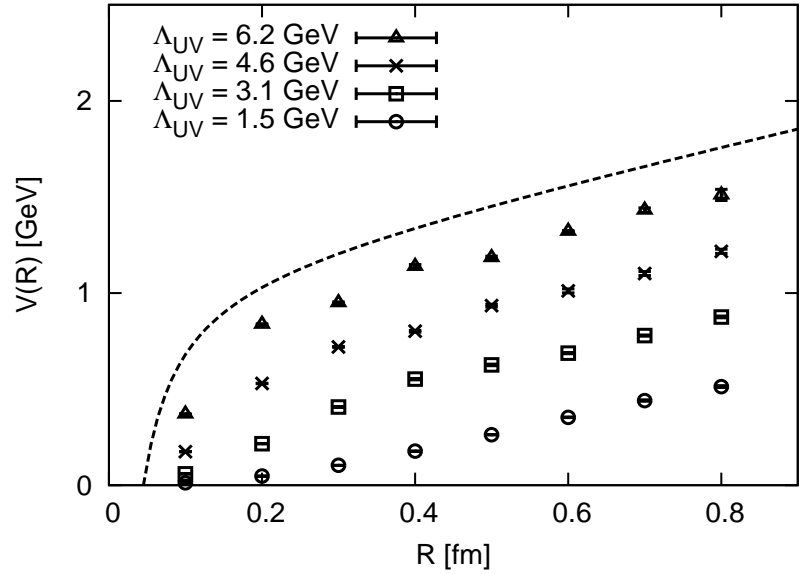}
\caption{\label{Fig3-1}
The static quark-antiquark potential $V(R)$ with the UV cutoff $\Lambda_{\rm UV}$.
The lattice QCD calculation is performed on $16^4$ lattice with $\beta =6.0$.
The broken line is the original quark-antiquark potential.
}
\end{center}
\end{figure}

We show the quark-antiquark potential with the UV cutoff in Fig.~\ref{Fig3-1}.
When the high-momentum gluon is removed by the UV cutoff, the short-range Coulomb potential is gradually reduced.
This is because the short-range Coulomb potential is an UV property.
The constant term $C$ is also reduced by the UV cutoff.
The constant term is mainly given by the lattice regularization for the UV singularity.
On the other hand, the long-range linear confinement potential is almost unaffected by the UV cutoff.
At $\Lambda_{\rm UV}=1.5$ GeV, the quark-antiquark potential becomes the linear potential in the whole region.
Note that the quark-antiquark potential would include $-\pi / (12R)$ term, so-called the L\"{u}scher term \cite{Lu80,Lu81}.
The L\"{u}scher term is similar to the perturbative Coulomb potential, but its origin is the nonperturbative string fluctuation.
It is nontrivial whether or not the long-range L\"{u}scher term remains in this calculation.
To reveal this, one needs to carefully analyze the functional form of the potential in long range with high precision.

\begin{figure}[p]
\begin{center}
\includegraphics[scale=1.5]{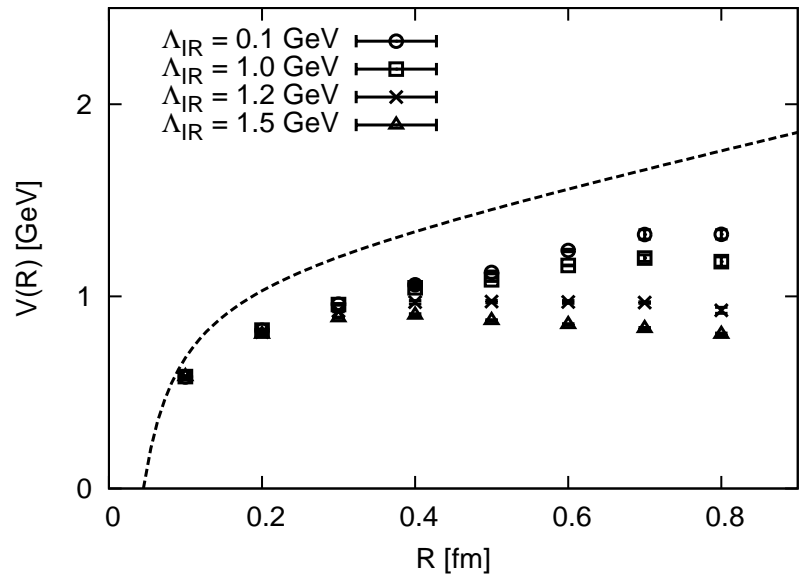}
\caption{\label{Fig3-2}
The static quark-antiquark potential $V(R)$ with the IR cutoff $\Lambda_{\rm IR}$.
}
\end{center}
\end{figure}

We show the quark-antiquark potential with the IR cutoff in Fig.~\ref{Fig3-2}.
The linear confinement potential is strongly affected by the IR cutoff.
At $\Lambda_{\rm IR}=0.1$ GeV and 1.0 GeV, the slope of the linear confinement potential, i.e., the string tension, is reduced from the original one.
At $\Lambda_{\rm IR}= 1.5$ GeV, the linear confinement potential completely disappears.
The potential becomes the Coulomb-like form.

From these results, we conclude as follows.
The perturbative and nonperturbative parts of the quark-antiquark potential are decoupled in the momentum space of the gluon.
The low-momentum gluon induces the confinement potential, and the high-momentum gluon induces the Coulomb potential.

\begin{figure}[p]
\begin{center}
\includegraphics[scale=1.5]{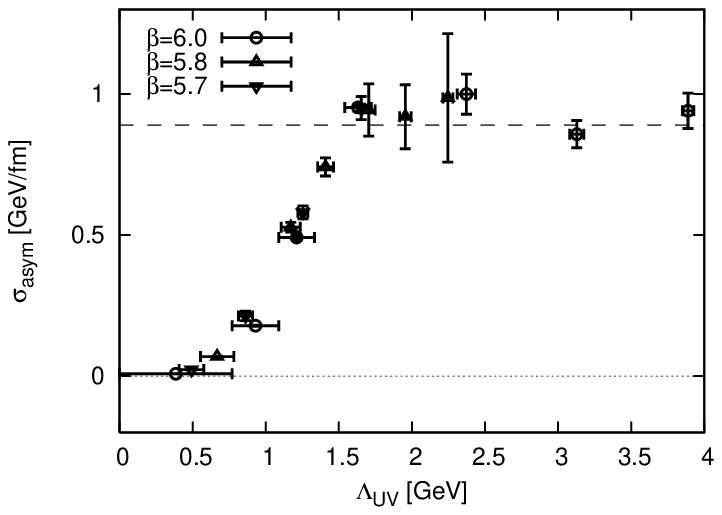}
\caption{\label{Fig3-3}
The asymptotic string tension $\sigma_{\rm asym}$ with the UV cutoff $\Lambda_{\rm UV}$.
The vertical error bar is the standard statistical error, and the horizontal error bar is the range that yields the same result due to the discrete momentum.
The original value of the string tension is $\sigma \simeq 0.89$ GeV/fm (broken line).
}
\end{center}
\end{figure}

\begin{figure}[p]
\begin{center}
\includegraphics[scale=1.5]{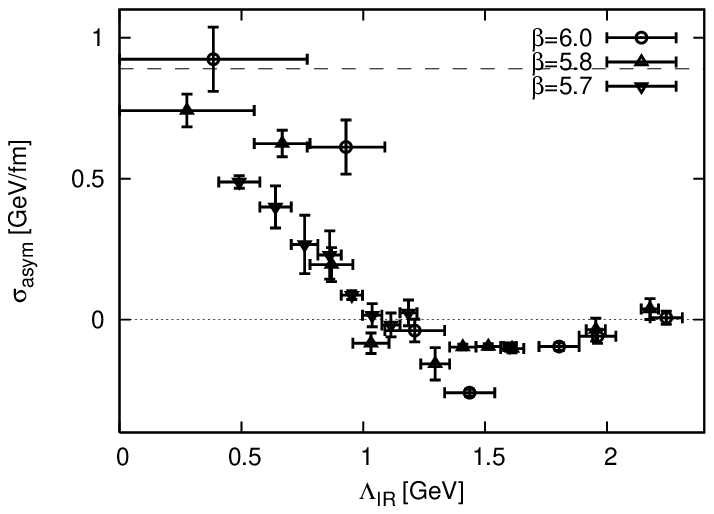}
\caption{\label{Fig3-4}
The asymptotic string tension $\sigma_{\rm asym}$ with the IR cutoff $\Lambda_{\rm IR}$.
}
\end{center}
\end{figure}

Next, we quantitatively determine which energy-momentum component of the gluon field induces the confinement potential.
For a quantitative argument, we need higher accuracy in momentum space.
Since the minimum momentum on lattice is the momentum-space lattice spacing $a_p$, the four-momentum length is restricted to discrete values as
\begin{eqnarray}
\sqrt{p^2}=\sqrt{\sum_{\mu}p_\mu p_\mu}=0, a_p, \sqrt{2}a_p, \sqrt{3}a_p, \cdots.
\end{eqnarray}
We can only take discrete variation on the value of the momentum cutoff.
For example, the lattice calculations in the range $0<\Lambda_{\rm IR}\le a_p$ yield the same result.
This is, so to speak, a discretization error in momentum space.
In order to achieve a finer resolution in momentum space, we have to calculate with smaller momentum-space lattice spacing, i.e., larger coordinate-space lattice volume.

For this purpose, we calculate the quark-antiquark potential on $16^4$ lattice with different three lattice spacings.
As listed in Table \ref{tab3-1}, the corresponding momentum-space lattice spacings are $a_p\simeq 0.41$ GeV, 0.55 GeV, and 0.77 GeV.
We estimate the asymptotic string tension $\sigma_{\rm asym}$ by fitting the quark-antiquark potential with a linear function $\sigma_{\rm asym}R + {\rm const.}$ in $0.3 \ {\rm fm}< R < 0.9$ fm.

We show the asymptotic string tension with the UV cutoff in Fig.~\ref{Fig3-3} and with the IR cutoff in Fig.~\ref{Fig3-4}.
In these figures, the vertical error bar represents the standard statistical error, and the horizontal error bar is not the statistical error but the range which yields the same result due to the discrete momentum.
In Fig.~\ref{Fig3-3}, the asymptotic string tension is almost unaffected in $\Lambda_{\rm UV}> 1.5$ GeV, and its value is still its original value $\sigma \simeq 0.89$ GeV/fm.
In $\Lambda_{\rm UV}< 1.5$ GeV, the asymptotic string tension is drastically reduced.
In Fig.~\ref{Fig3-4}, the asymptotic string tension is sensitive to the IR cutoff.
In $\Lambda_{\rm IR}=1.5$ GeV, the asymptotic string tension is almost zero and the confinement potential disappears.

From these quantitative analyses, we conclude that the relevant energy scale of color confinement is below about 1.5 GeV, i.e., color confinement is induced by the low-momentum gluon below about 1.5 GeV.

\section{Other conditions}
We have calculated with other conditions for the check of consistency: a different gauge and the three-dimensional formalism.
Here, we briefly summarize these calculations and results.

\begin{table}[p]
\newcommand{\m}{\hphantom{$-$}}
\newcommand{\cc}[1]{\multicolumn{1}{c}{#1}}
\renewcommand{\tabcolsep}{1pc} 
\renewcommand{\arraystretch}{1} 
\caption{\label{tab3-2}
Asymptotic string tension $\sigma_{\rm asym}$ with the IR cutoff $\Lambda_{\rm IR}$.
The results obtained under three different conditions are listed; the main result with the four-dimensional momentum cutoff in the Landau gauge, the result in the Coulomb gauge, and the result with three-dimensional formalism in the Landau gauge.
The lattice QCD calculations are performed on $16^4$ lattice with $\beta =6.0$.
}
\begin{center}
\begin{tabular}{cccc}
\hline\hline
$\Lambda_{\rm IR}$ [GeV] &$\sigma_{\rm asym}a^2 (4-dim.)$ & $\sigma_{\rm asym}a^2$ (Coulomb)& $\sigma_{\rm asym}a^2$ (3-dim.)\\
\hline
0 & 0.051 & 0.051 & 0.051\\
$\sim$ 0.1 & 0.0469(58) & 0.0289(58) & 0.0433(51)\\
1.0 & 0.0311(49) & 0.0190(59) & 0.0198(43)\\
1.2 & -0.0019(20)& 0.0024(25) & -0.0034(9)\\
1.5 & -0.0132(6) & -0.0041(10)& -0.0092(5)\\
2.2 & 0.0003(12) & 0.0058(10) & 0.0024(13)\\
\hline\hline
\end{tabular}
\end{center}
\end{table}

\begin{figure}[p]
\begin{center}
\includegraphics[scale=1.5]{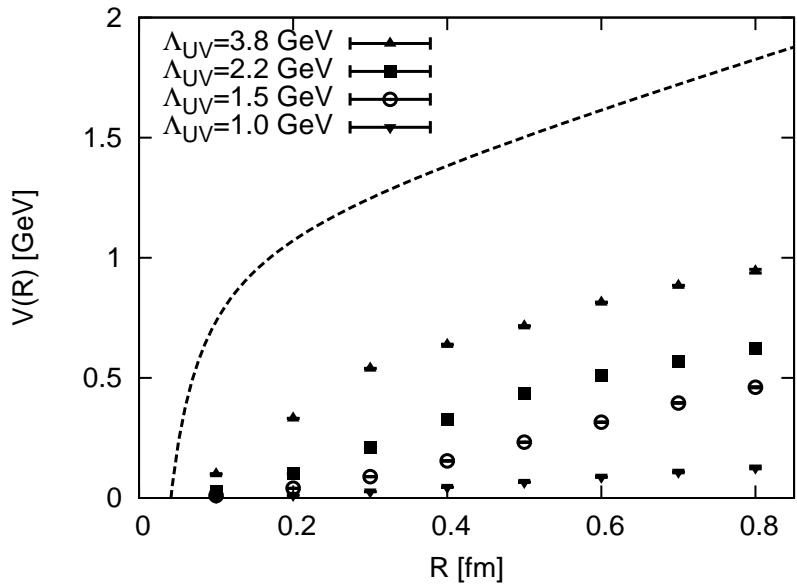}
\caption{\label{Fig3-5}
The quark-antiquark potential $V(R)$ with the UV cutoff $\Lambda_{\rm UV}$ in the three-dimensional formalism.
}
\end{center}
\end{figure}

\subsubsection{Different gauge}
Since our framework is not gauge invariant, it is important to take other gauge choices.
We calculate with the Coulomb gauge, instead of the Landau gauge.
The gauge-fixing condition for the Coulomb gauge is to maximize the quantity
\begin{eqnarray}
F_C[U] \equiv \sum_{x} \sum_{j=1}^{3} {\rm ReTr} U_j (x).
\end{eqnarray}
The fitting results of $\sigma_{\rm asym}$ are listed in the third column of Table \ref{tab3-2}.
The values itself are different, but the qualitative behavior is the same as that of the Landau gauge.

Which energy-momentum component induces the confinement potential, in principle, depends on the gauge choice.
In the present framework, we can calculate with any other gauges in the same way.

\subsubsection{Three-dimensional formalism}
Our framework can be easily extended to the spatial three-dimensional formalism.
We perform the three-dimensional Fourier transformation and consider the cutoff by the three-momentum length $|\vec{p}|$.
The results are listed in the fourth column of Table \ref{tab3-2}.
The behavior is almost the same as the case of the four-dimensional momentum cutoff.
The quark-antiquark potential with the UV cutoff $\Lambda_{\rm UV}$ is shown in Fig.~\ref{Fig3-5}.
The confinement potential is not affected in $\Lambda_{\rm UV}>1.5$ GeV, and is affected in $\Lambda_{\rm UV}<1.5$ GeV.
Thus, the confinement potential is induced by the low-momentum gluon below about 1.5 GeV also in the three-dimensional formalism.

\section{Comparison with an analytical model}
In this section, we analyze the Richardson potential, which is a phenomenological model of the interquark potential \cite{Ri79}.
We introduce the momentum cutoff to the Richardson potential and compare it with the lattice QCD result.

The Richardson potential is constructed so as to reproduce the Coulomb plus linear functional structure.
It is defined by the one-dressed-gluon-exchange amplitude which is proportional to 
\begin{eqnarray}
\tilde{V}(p^2)=-C_F\frac{{\bar g}^2(p^2)}{p^2},
\end{eqnarray}
where
\begin{eqnarray}
{\bar g}^2(p^2)=\frac{1}{\beta_0 \ln(1+p^2/\Lambda^2)}.
\end{eqnarray}
Here, $C_F=4/3$ and $\beta_0=(11N_c-2N_f)/48\pi^2$.
$\Lambda$ is a parameter in this model.
This coupling is similar to the standard QCD coupling, except for 1 in the argument of logarithm.
To obtain the nonrelativistic (instantaneous) potential, we calculate the three-dimensional Fourier transformation as
\begin{eqnarray}
V(R)&=&\int \frac{d^3p}{(2\pi)^3}\ e^{i\vec{R}\cdot \vec{p}}\ \tilde{V}(\vec{p}^2)\nonumber\\
&=&\frac{C_{\rm F}}{8\pi \beta_0}\left[ \Lambda^2 R - \frac{1}{R} + \frac{f(\Lambda R)}{R} \right],
\label{VRi}
\end{eqnarray}
where
\begin{eqnarray}
f(x)=4\int_{1}^{\infty}dt \frac{e^{-tx}}{t}\frac{1}{[\ln(t^2-1)]^2+\pi^2}.
\end{eqnarray}
We set $N_f=0$ to compare this potential with the quark-antiquark potential in quenched lattice QCD, and $\Lambda=0.48$ GeV so that the coefficient of the linear potential is equal to the physical string tension $\sigma \simeq 0.89$ GeV/fm.

We can consider two types of the IR cutoff.
One is the simple IR cutoff by the three-momentum length on the Fourier transformation,
\begin{eqnarray}
V(R)=\int_{|\vec{p}|\ge \Lambda_{\rm IR}} \frac{d^3p}{(2\pi)^3} \ e^{i\vec{R}\cdot \vec{p}}\ \tilde{V}(\vec{p}^2).
\end{eqnarray}
The other is the change of the functional form as
\begin{eqnarray}
\tilde{V}(\vec{p}^2)=-C_F\frac{{\bar g}^2(\vec{p}^2)}{\vec{p}^2+\Lambda_{\rm IR}^2}.
\label{VRiIR1}
\end{eqnarray}
The advantage of the latter way is that we can analytically calculate the momentum integral.
The result is
\begin{eqnarray}
V(R)=\frac{C_{\rm F}}{8\pi \beta_0} \left[ -\frac{2}{R} \Big(\frac{1}{\lambda^2}+h_\lambda e^{-\lambda \Lambda R}\Big) + \frac{f_\lambda (\Lambda R)}{R} \right], 
\label{VRiIR2}
\end{eqnarray}
where
\begin{eqnarray}
h_\lambda &=& \Bigg\{
\begin{array}{cc}
\frac{\ln (\lambda^2-1)}{[\ln (\lambda^2-1)]^2+\pi^2} & (\lambda >1)\\
\frac{1}{\ln (1- \lambda^2)} & (1 \ge\lambda\ge 0)
\label{hl}
\end{array}\\
f_\lambda(x)&=&4\mathcal{P}\int_{1}^{\infty}dt \ \frac{te^{-tx}}{t^2-\lambda^2} \frac{1}{[\ln(t^2-1)]^2+\pi^2}
\label{fl}
\end{eqnarray}
and $\lambda=\Lambda_{\rm IR}/\Lambda$.
The symbol $\mathcal{P}$ indicates the principal value of the integral, which is necessary in the case of $\lambda>1$.
This functional form becomes the original Richardson potential in the limit of $\Lambda_{\rm IR}\to 0$, apart from an irrelevant constant.

\begin{figure}[t]
\begin{center}
\includegraphics[scale=1.2]{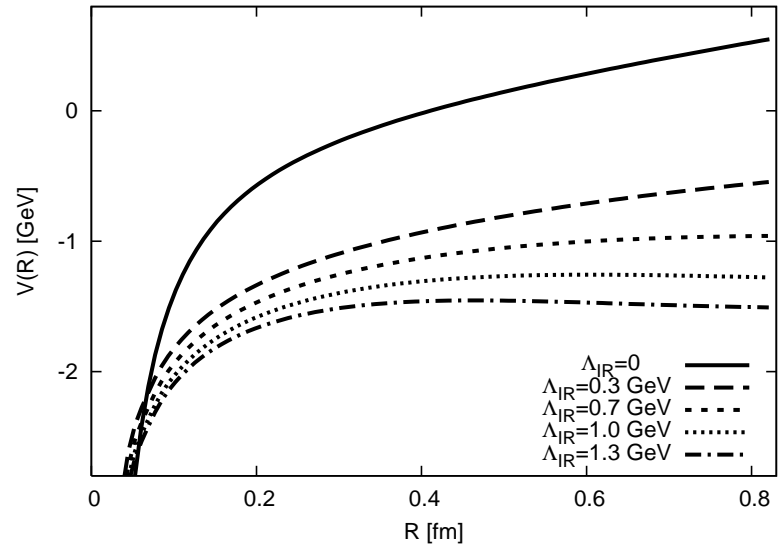}
\caption{\label{Fig3-6}
The Richardson potential with the IR cutoff $\Lambda_{\rm IR}$.
The solid line is the original Richardson potential (\ref{VRi}).
The irrelevant constant is arbitrarily subtracted.
}
\end{center}
\end{figure}

These two ways to introduce the IR cutoff yield almost the same results.
The result of the former way is shown in Fig.~\ref{Fig3-6}.
The irrelevant constant is arbitrarily subtracted in the figure.
As in the case of lattice QCD, the string tension decreases when the IR cutoff is introduced.
In $\Lambda_{\rm IR}> 1$ GeV, the linear confinement potential disappears and the interquark potential becomes a Coulomb-like potential.
This behavior is consistent with the lattice QCD result.
While the Richardson potential is only a phenomenological model and its confinement potential is set by hand, it well reproduces the $\Lambda_{\rm IR}$-dependence of the interquark potential obtained by lattice QCD.

\chapter{Color Flux Tube}
\thispagestyle{headings}
In a meson, the gluon forms a string-like structure, the so-called color flux tube, between the quark and the antiquark.
The color flux tube is considered to be essential for color confinement.
Although the color flux tube is not directly observed in experiments, it can be observed in lattice QCD.
In this chapter, we apply our framework to the analysis of the color flux tube.

The color flux tube is measured by calculating the action density distribution around a static quark-antiquark pair \cite{So87,Ha87}.
At the positions of color sources, the action density distribution has the peaks which is perturbative singularities of the self energies.
The color flux tube is formed between the peaks of color sources, but its contribution is small compared to perturbative contributions.
In Chapter 3, we found that the confinement potential originates from the IR gluon below 1.5 GeV in the Landau gauge.
Thus, if we remove the high-momentum component above 1.5 GeV by the present framework, we can extract the color flux tube from the action density distribution.
In other words, we can clearly observe the structure of the color flux tube apart from unnecessary perturbative contributions.
In this chapter, we consider the three-dimensional UV cutoff explained in Section 3.2.

\begin{table}[h]
\renewcommand{\tabcolsep}{1pc} 
\renewcommand{\arraystretch}{1} 
\caption{\label{tab4-1}
Simulation parameters of the gauge configurations.
The notation is the same as in Table \ref{tab3-1}.
}
\begin{center}
\begin{tabular}{cccccc}
\hline\hline
& $\beta$ & $V$ [$a^4$] & $a$ [fm] & $a_p$ [GeV] & $N_{\rm conf}$\\
\hline
Quenched & 6.0 & $16^4$ & 0.10 & 0.77 & 500\\
Quenched & 6.0 & $32 \times 16^3$ & 0.10 & 0.77 & 500\\
\hline\hline
\end{tabular}
\end{center}
\end{table}

\section{Action density in vacuum}

Before the action density distribution with a static quark-antiquark pair, we calculated the action density without color sources, i.e., in vacuum.
The action density $s (x)$ is defined as Eq.~(\ref{eqAD}).
The vacuum action density relates to the gluon condensate in continuum QCD, which produces the trace anomaly.
In the naive continuum limit, the action density corresponds to the gluon condensate, however, at finite lattice spacing, it is dominated by perturbative corrections \cite{Ba81,Gi81}.
In short, a large part of the vacuum action density on the lattice is perturbatively generated.

The vacuum action density $\langle s (x) \rangle$ with the UV cutoff is listed in Table \ref{tab4-2}.
The vacuum action density is translationally invariant, i.e., independent of $x$.
When the high-momentum gluon is removed, the vacuum action density drastically decreases.
For example, at $\Lambda_{\rm UV}= 1.5$ GeV, the vacuum action density is reduced to about 1 \% of the original value.
The remaining small component would lead to nonperturbative properties of QCD vacuum.

\begin{table}[h]
\renewcommand{\tabcolsep}{1pc} 
\renewcommand{\arraystretch}{1} 
\caption{\label{tab4-2}
The UV cutoff $\Lambda_{\rm UV}$, the asymptotic string tension $\sigma_{\rm asym}$ of the quark-antiquark potential, and the vacuum action density $\langle s (x) \rangle$.
The statistical error of $\langle s (x) \rangle$ is omitted because it is negligibly small.
}
\begin{center}
\begin{tabular}{ccccc}
\hline\hline
$\Lambda_{\rm UV}$ [GeV] &$\sigma_{\rm asym}$ [GeV/fm] & $\langle s (x) \rangle$ [$a^{-4}$] \\
\hline
No Cut & 0.89 & 14.51 \\
3.8 & 0.824(31) & 2.57 \\
2.2 & 0.801(67)& 0.60 \\
1.5 & 0.799(18) & 0.20\\
1.0 & 0.208(4) & 9.5$\times 10^{-3}$\\
\hline\hline
\end{tabular}
\end{center}
\end{table}

\section{Action density with a quark-antiquark pair}

\begin{figure}[p]
\begin{center}
\includegraphics[scale=0.5]{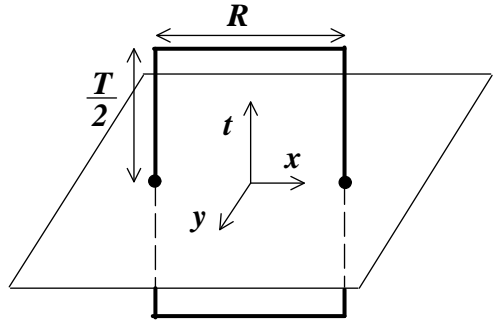}
\caption{\label{Fig4-1}
The Wilson loop $W(R,T)$ and the three-dimensional plane where the action density $\langle s (x) \rangle_W$ is measured.
The origin of the coordinate is placed in the center of the Wilson loop. 
}
\end{center}
\end{figure}

The spatial distribution of the action density around a static quark-antiquark pair is obtained by measuring $s (x)$ around the Wilson loop at a certain time slice.
Its expectation value is given by
\begin{eqnarray}
\langle s (x) \rangle_W \equiv \frac{\langle s (x) W(R,T) \rangle}{\langle W(R,T) \rangle} - \langle s (x) \rangle,
\label{rhoW}
\end{eqnarray}
where $W(R,T)$ is the value of the Wilson loop with the size $R \times T$.
The schematic figure is illustrated in Fig.~\ref{Fig4-1}.
The origin of the four-dimensional coordinate is placed in the center of the Wilson loop. 
We measured $\langle s (x) \rangle_W$ at the $t=0$ plane.

For statistical improvement, the translationally invariant quantities are averaged over, if possible.
For example, $\langle \rho (x) W(R,T) \rangle$ is given by the convolution sum as
\begin{eqnarray}
\langle \rho (x) W(R,T) \rangle = \langle \frac{1}{V}\sum_{s} \rho (x+s) W(R,T,s) \rangle,
\end{eqnarray}
where $s$ is the position of the parallel-translated Wilson loop.
To enhance the ground-state component, the APE smearing method is applied to the spatial link variables of the Wilson loop \cite{Al87}.
The action density distribution $\langle s (x) \rangle_W$ is independent of the time slice if the ground-state component is suitably dominated.
Because of the smearing method, the action density distribution is almost independent of $T$ in the range of $T \ge 4a$ in the present calculation.

We introduce the UV cutoff to the action density distribution as
\begin{eqnarray}
\label{eqADUV}
\langle s [U^\Lambda] \rangle_W \equiv \frac{\langle s [U^\Lambda] W[U] \rangle}{\langle W[U] \rangle} - \langle s [U^\Lambda] \rangle.
\end{eqnarray}
The arguments, such as $x$, are abbreviated for simplicity.
The physical interpretation is the spatial distribution of the low-momentum gluon around a physical quark-antiquark pair.
As another choice, one can introduce the UV cutoff not only for $s [U]$ but also for $W[U]$, and its result is expected to be qualitatively similar to Eq.~(\ref{eqADUV}).

\begin{figure}[p]
\begin{center}
\includegraphics[scale=1.8]{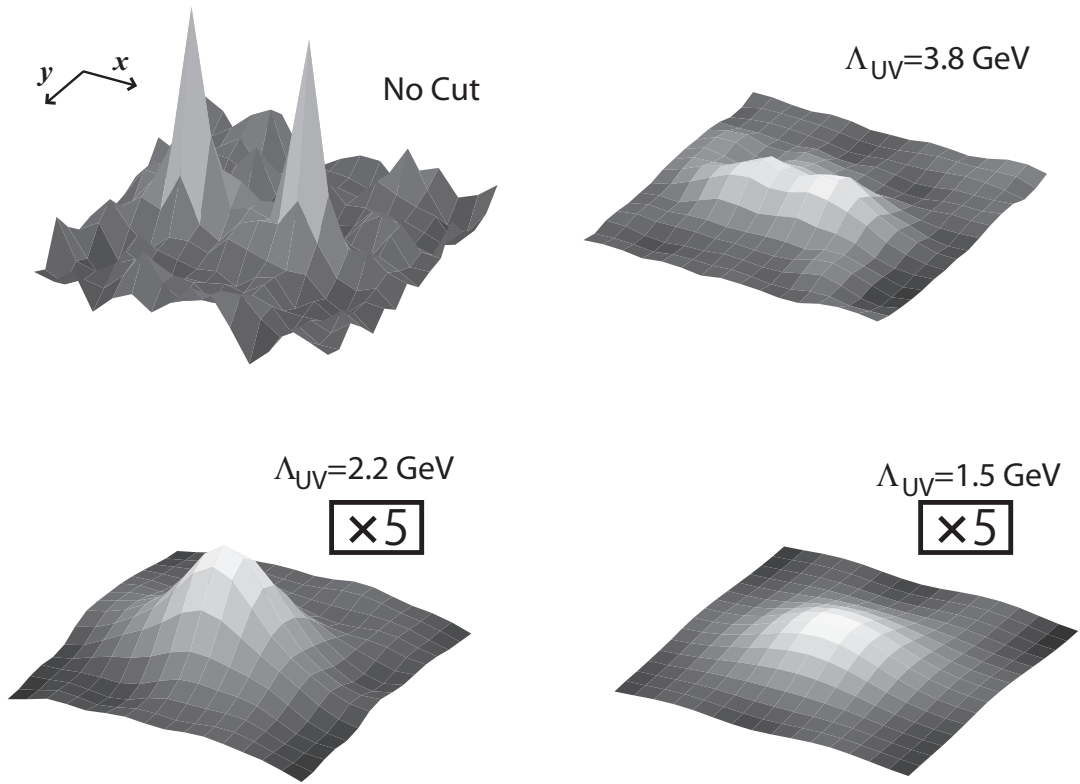}
\caption{\label{Fig4-2}
The action density distribution $\langle s (x) \rangle_W$ around a static quark-antiquark pair.
The separation between the quark and antiquark is $R=0.6$ fm.
``$\times 5$''s mean that the lower figures are five times enlarged in the vertical direction compared to the upper figures.
The ``No Cut'' is the original lattice QCD result.
}
\end{center}
\end{figure}

In Fig.~\ref{Fig4-2}, we display the action density distribution $\langle s (x) \rangle_W$ with the UV cutoff $\Lambda_{\rm UV}=1.5$ GeV, 2.2 GeV, and 3.8 GeV.
The interquark distance between the quark and the antiquark is $R=0.6$ fm.
The overall sign of $\langle s (x) \rangle_W$ is flipped in the figure, which is only a matter of definition.
Because the absolute value of $\langle s (x) \rangle_W$ at $\Lambda_{\rm UV}=1.5$ GeV and 2.2 GeV is small, these data in Fig.~\ref{Fig4-2} are enlarged by a factor of five compared to the other ones.
The action density distribution in original lattice QCD without the momentum cutoff is also displayed (``No Cut'' in the upper left); however, its statistical error is relatively large.
In original lattice QCD, the action density is strongly enhanced at the positions of the quark and the antiquark.
In contrast, the color flux tube is difficult to observe because of such singular peaks and the large statistical fluctuation.
Both the singular peaks and the large fluctuation originate from the perturbative part of the action density.
When the high-momentum gluon above 3.8 GeV is removed (the upper right), these perturbative contributions are drastically suppressed, and the flux-tube structure connecting the quark and the antiquark becomes clear.
At $\Lambda_{\rm UV}=1.5$ GeV and 2.2 GeV (the lower right and the lower left, respectively), the two peaks  seem to disappear, and the action density is distributed around the origin.

Apart from the vacuum contribution, which is translationally invariant, the action density at $\Lambda_{\rm UV}=1.5$ GeV is broadly distributed around the midpoint between the quark and the antiquark.
In the calculation of the quark-antiquark potential in Chapter 3, this low-momentum gluon leads to the linear confinement potential over the entire range of $R$.
Therefore, this action density distribution corresponds to the color flux tube inducing the confinement potential.

\begin{figure}[p]
\begin{center}
\includegraphics[scale=1.5]{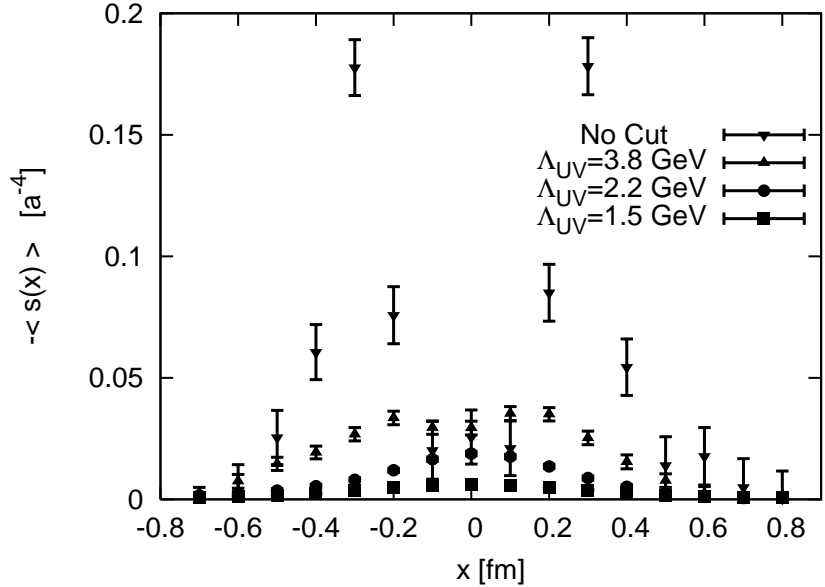}
\caption{\label{Fig4-3}
The action density distribution $\langle s (x) \rangle_W$ along the $x$-axis.
The interquark distance $R$ is 0.6 fm. 
The quark and the antiquark are located on $x=0.3$ fm and $x=-0.3$ fm, respectively.
The ``No Cut'' is the original lattice QCD result.
}
\end{center}
\end{figure}

\begin{figure}[p]
\begin{center}
\includegraphics[scale=1.5]{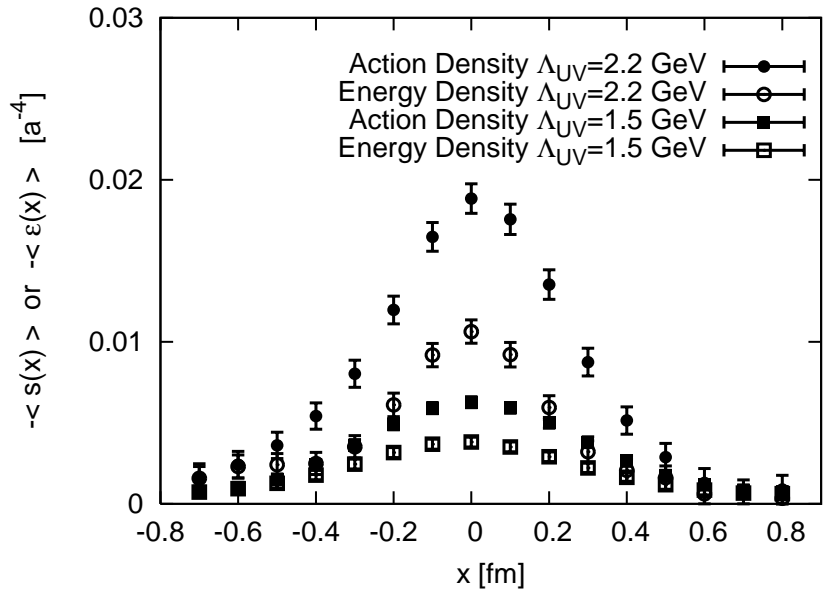}
\caption{\label{Fig4-4}
The action density distribution $\langle s (x) \rangle_W$ and the energy density distribution $\langle \varepsilon (x) \rangle_W$ along the $x$-axis.
The interquark distance $R$ is 0.6 fm. The quark and the antiquark are located on $x=0.3$ fm and $x=-0.3$ fm, respectively.
}
\end{center}
\end{figure}

In Fig.~\ref{Fig4-3}, we plot the action density distribution $\langle s (x) \rangle_W$ along the $x$-axis, i.e., in the longitudinal direction of the quark-antiquark separation.
In original lattice QCD without the UV cutoff (``No Cut''), the action density has two self-energy peaks at the positions of the quark and the antiquark, $x=0.3$ fm and $x=-0.3$ fm, respectively.
When the UV cutoff is introduced, these self-energy peaks are drastically suppressed.
Moreover, the absolute values of the action density and the statistical fluctuation become small, as the vacuum action density.
The results at $\Lambda_{\rm UV}=1.5$ GeV and 2.2 GeV are also shown in Fig.~\ref{Fig4-4}.
The action density distribution has a maximum at the origin, and the self-energy peaks seem to disappear.
Although the self-energy peaks would also include a nonperturbative contribution, it is too small to distinguish from the flux tube.
The endpoints of the flux tube are not sharp, spreading outside the positions of the quark and the antiquark.
In the case of $R=0.6$ fm, the longitudinal shape of the flux tube resembles a broad mountain rather than a plateau.

We also show the energy density distribution $\langle \varepsilon (x) \rangle_W$ along the $x$-axis in Fig.~\ref{Fig4-4}.
The energy density $\varepsilon (x)$ is defined identically as the action density by changing the relative sign between the spatial plaquettes and the temporal plaquettes.
The absolute value of the energy density is smaller than that of the action density due to the cancellation between the chromoelectric contribution and the chromomagnetic contribution.
Apart from the absolute value, the overall shape of the energy density distribution is similar to that of the action density distribution.

\begin{figure}[p]
\begin{center}
\includegraphics[scale=1.5]{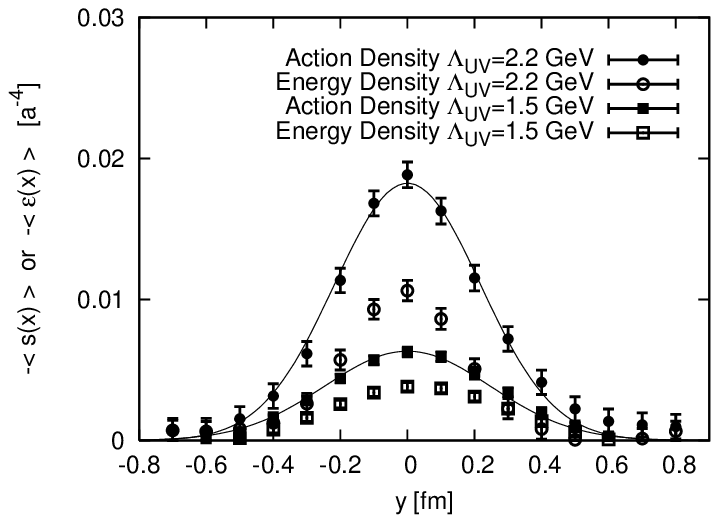}
\caption{\label{Fig4-5}
The action density distribution $\langle s (x) \rangle_W$ and the energy density distribution $\langle \varepsilon (x) \rangle_W$ along the $y$-axis.
The interquark distance $R$ is 0.6 fm. 
The solid lines are the results of fitting to a the Gaussian function $s_0 \exp (-y^2/\delta^2)$.
}
\end{center}
\end{figure}

To estimate the width of the flux tube, we fit the action density
distribution along the $y$-axis to a the Gaussian form, $s_0 \exp (-y^2/\delta^2)$.
It is seen that the Gaussian form can well reproduce the lattice data, as shown in Fig.~\ref{Fig4-5}.
The best-fit width parameter $\delta$ is 0.31$\pm$0.01 fm at $\Lambda_{\rm UV}=2.2$ GeV, and 0.35$\pm$0.01 fm at $\Lambda_{\rm UV}=1.5$ GeV.
These values are comparable to the flux-tube width of earlier works in
the standard lattice QCD \cite{Gi90,Ba95,Ha96,Pe97}.

\begin{figure}[pj]
\begin{center}
\includegraphics[scale=2.4]{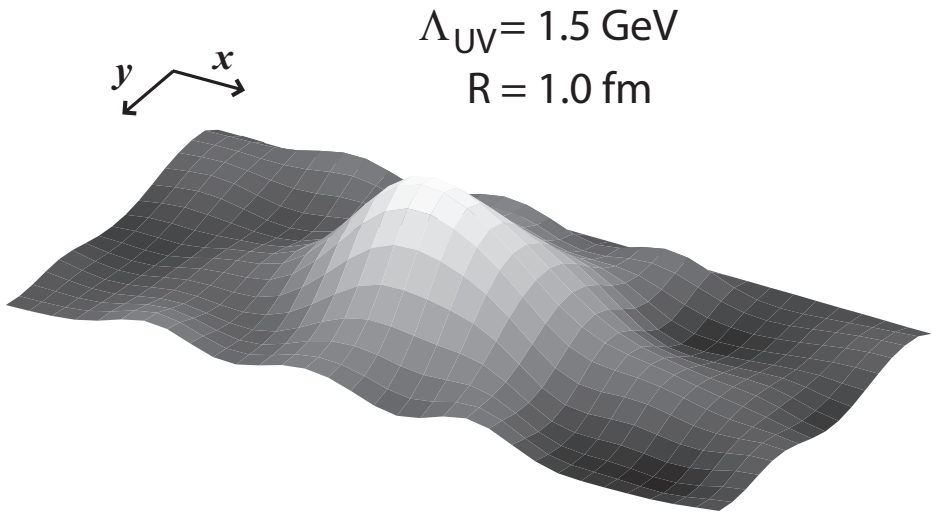}
\caption{\label{Fig4-6}
The action density distribution $\langle s (x) \rangle_W$ for the interquark distance $R=1.0$ fm.
This calculation is performed on a 32$\times 16^3$ lattice.
}
\end{center}
\end{figure}

Next, we analyze how the flux-tube shape depends on the interquark distance $R$.
As shown in Fig.~\ref{Fig4-2}, the longitudinal length and the transverse width are almost the same at $R=0.6$ fm and $\Lambda_{\rm UV}=1.5$ GeV.
The overall shape seems to be isotropic and far from a ``string'' or ``tube''.
This is because the transverse width of the flux tube is fairly large.
To approach the distribution to the tube-like shape, the interquark distance must be enlarged \cite{Ba95,Ha96,Pe97}.
Here, we use a 32($x$-axis)$\times 16^3$ lattice instead of a $16^4$ lattice and extend the interquark distance to $R=1.0$ fm.
As shown in Fig.~\ref{Fig4-6}, the distribution is stretched in the longitudinal direction, and the longitudinal length becomes larger than the transverse width.
Compared to the result of $R=0.6$ fm, the flux tube at $R=1.0$ fm approaches a broad tube-like shape.

\chapter{Meson Mass}
\thispagestyle{headings}
In this chapter, we calculate the mass of a pion (pseudoscalar meson) and a $\rho$-meson (vector meson) in quenched lattice QCD.
The physical values are $m_\pi \simeq$ 140 MeV and $m_\rho \simeq$ 770 MeV.
The ground-state mass of a meson is extracted from the large-time behavior of the corresponding meson correlator as
\begin{eqnarray}
\langle M_\Gamma^\dagger(\vec{x},t) M_\Gamma(\vec{x},0) \rangle = C_0 e^{- m_\Gamma t} + \cdots.
\end{eqnarray}
The interpolation operator $M_\Gamma(\vec{x},t)$ is, for example, $\bar{q}\gamma_5 q$ for a pion and $\bar{q}\gamma_j q$ for a $\rho$-meson.
The dots contain the excited-state components, which rapidly die out in large $t$.

We used the clover fermion and the staggered fermion for the quark propagator.
The clover fermion is the $O(a)$-improved Wilson fermion, and the quark mass is described in terms of the mean-field-improved hopping parameter $\kappa$ \cite{Sh85,Le93}.
The setups of the gauge configurations are shown in Table \ref{tab5-1}.
The clover fermion and the staggered fermion are used on $16^4$ and $16^3 \times 32$ lattices, respectively.

\begin{table}[h]
\renewcommand{\tabcolsep}{1pc} 
\renewcommand{\arraystretch}{1} 
\caption{\label{tab5-1}
Simulation parameters of the gauge configurations.
The notation is the same as in Table \ref{tab3-1}.
}
\begin{center}
\begin{tabular}{cccccc}
\hline\hline
& $\beta$ & $V$ [$a^4$] & $a$ [fm] & $a_p$ [GeV] & $N_{\rm conf}$\\
\hline
Quenched & 6.0 & $16^4$ & 0.10 & 0.77 & 100\\
Quenched & 6.0 & $16^3 \times 32$ & 0.10 & 0.77 & 100\\
\hline\hline
\end{tabular}
\end{center}
\end{table}

\section{Meson mass with the momentum cutoff}
\begin{figure}[p]
\begin{center}
\includegraphics[scale=1.5]{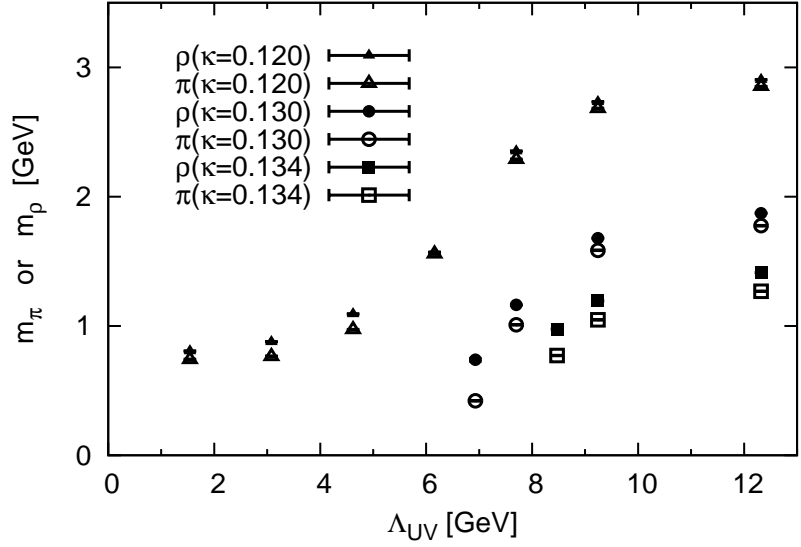}
\caption{\label{fig5-1}
The $\Lambda_{\rm UV}$-dependence of pion mass $m_\pi$ and $\rho$-meson mass $m_\rho$ with the clover fermion.
The data at $\Lambda_{\rm UV}\simeq 12.5$ GeV is the original lattice result.
}
\end{center}
\end{figure}

In Fig.~\ref{fig5-1}, we show the meson masses with the UV cutoff with the clover fermion.
The data at $\Lambda_{\rm UV}\simeq 12.5$ GeV are the original lattice results.
As the high-momentum components are removed, both the pion mass $m_\pi$ and the $\rho$-meson mass $m_\rho$ gradually decreases.
This is because the UV gluon ``dresses'' the quarks in the mesons.

\begin{figure}[p]
\begin{center}
\includegraphics[scale=1.5]{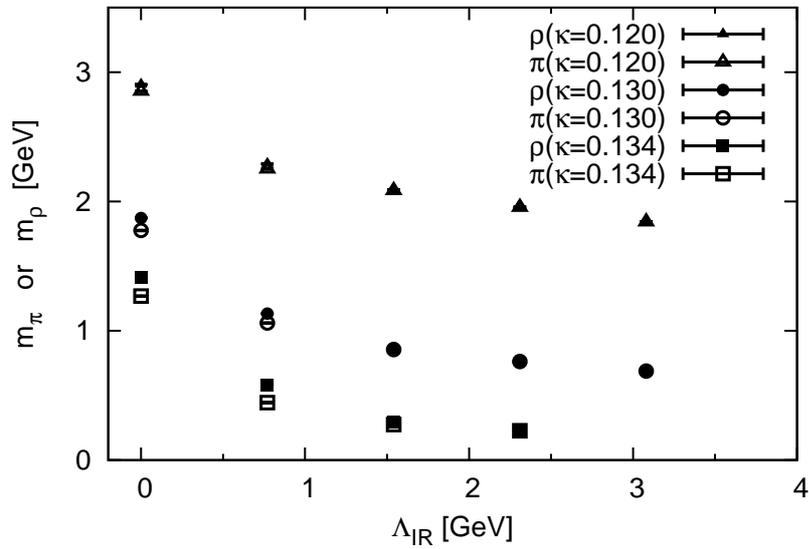}
\caption{\label{fig5-2}
The $\Lambda_{\rm IR}$-dependence of meson masses with the clover fermion.
The data at $\Lambda_{\rm IR}=0$ GeV is the original lattice result.
}
\end{center}
\end{figure}

In Fig.~\ref{fig5-2}, we show the meson masses with the IR cutoff with the clover fermion.
As in the case of the UV cutoff, the meson masses are reduced by the IR cutoff.
In addition, the pion mass and the $\rho$-meson mass degenerate in $\Lambda_{\rm IR}> 1.5$ GeV.
In Chapter 3, we found that the confinement potential disappears in this region.
Thus, this degeneracy suggests that the quark and antiquark become unbound or, if possible, very narrowly bound.
In such a state, the character of each meson is lost, and the meson mass is equal to the sum of the quark and antiquark masses.
This state is called as ``quasi-free''.

\begin{figure}[p]
\begin{center}
\includegraphics[scale=1.5]{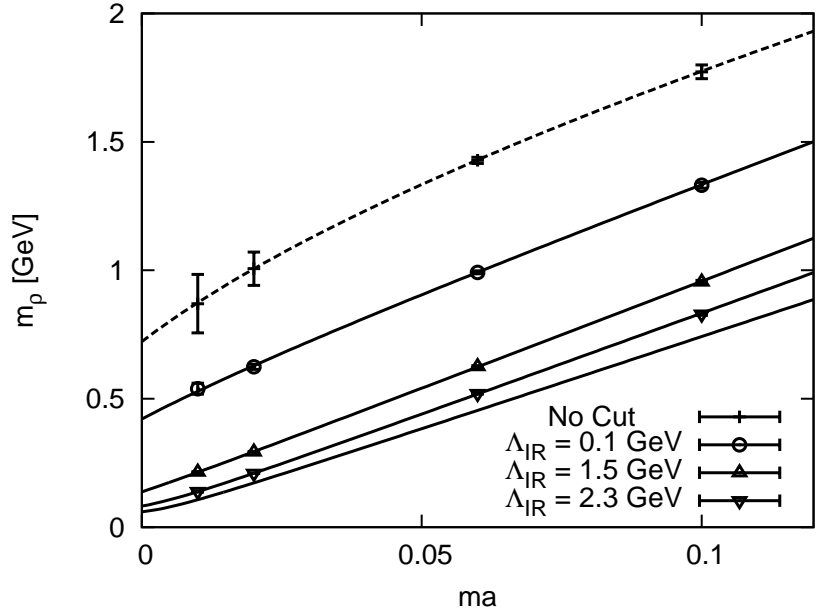}
\caption{\label{fig5-3}
The rho mass $m_\rho$ plotted against the bare quark mass $m$ with the staggered fermion.
The ``No Cut'' (broken line) is the original lattice QCD result.
}
\end{center}
\end{figure}

\begin{figure}[p]
\begin{center}
\includegraphics[scale=1.5]{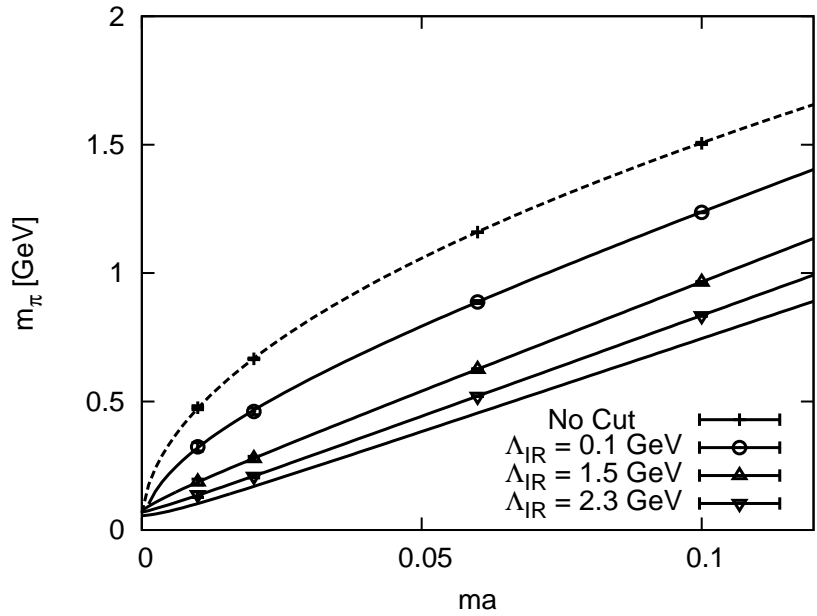}
\caption{\label{fig5-4}
The pion mass $m_\pi$ plotted against the bare quark mass $m$ with the staggered fermion.
}
\end{center}
\end{figure}

Next, we consider the dependence of the meson masses on the quark mass with the staggered fermion.
For chiral extrapolation, the squared meson masses are fitted with quadratic functions as
\begin{eqnarray}
 (m_\pi a)^2 \ {\rm or} \ (m_\rho a)^2 = c_2 (ma)^2 + c_1(ma) +c_0,
\end{eqnarray}
where $c_2$, $c_1$, and $c_0$ are fitting parameters. 
We show the results of chiral extrapolation in Fig.~\ref{fig5-3} and Fig.~\ref{fig5-4}.
When the IR cutoff is introduced, the $\rho$-meson mass uniformly decreases but the linear extrapolation form is unchanged.
In contrast, the functional form of the pion mass changes in $\Lambda_{\rm IR} \ge 1.5$ GeV, where the pion mass degenerates the $\rho$-meson mass.
The extrapolation function becomes a linear function.

The pion mass is related to the chiral condensate through the Gell-Mann-Oaks-Renner relation,
\begin{eqnarray}
f_\pi^2 m_\pi ^2 = - m \langle \bar{q}q \rangle .
\end{eqnarray}
In principle, we can examine spontaneous chiral symmetry breaking through the behavior of the pion mass.
However, especially when the quarks are deconfined, we have to take care of the finite-volume artifact.
We directly analyze spontaneous chiral symmetry breaking by calculating the chiral condensate in the next chapter.

\chapter{Chiral Condensate}
\thispagestyle{headings}

In this chapter, we analyze the chiral condensate.
The chiral condensate is an order parameter of spontaneous chiral symmetry breaking in the chiral limit.
It is nonzero in the symmetry-broken phase and zero in the symmetry-restored phase.
When the quark mass is finite, the chiral condensate includes the effect of explicit breaking by the quark mass as well as the dynamical effect.
To extract the chiral limit in lattice QCD, one calculates with several quark masses and extrapolates to the chiral limit.
We denote the absolute value of the chiral condensate as
\begin{eqnarray}
\Sigma \equiv - a^3 \langle \bar{q}q \rangle = a^3 {\rm tr} S_q,
\end{eqnarray}
where $S_q$ is the quark propagator.

We compute the chiral condensate with the staggered fermion in quenched QCD and in full QCD.
The full QCD configuration includes the two-flavor staggered fermion.
In full QCD, we use a single mass for the valence and sea quarks, $ma=m_{\rm sea}a=0.01$, of which the corresponding pion mass is about 500 MeV.
The simulation parameters of gauge configurations are summarized in Table \ref{tab6-1}.

\begin{table}[h]
\renewcommand{\tabcolsep}{0.9pc} 
\renewcommand{\arraystretch}{1} 
\caption{\label{tab6-1}
Simulation parameters of the gauge configurations.
The notation is the same as in Table \ref{tab3-1}.
The full QCD configuration is provided in the NERSC archive \cite{Br91}.
$m_{\rm sea}$ is the bare quark mass of the staggered fermion.
}
\begin{center}
\begin{tabular}{ccccccc}
\hline\hline
& $\beta$ & $m_{\rm sea}$ [$a^{-1}$] & $V$ [$a^4$] & $a$ [fm] & $a_p$ [GeV] & $N_{\rm conf}$\\
\hline
Full &  5.7 & 0.01 & $16^3\times 32$ & 0.10 & 0.79 & 24 - 49 \\
Quenched &  6.0 & - & $32^4$ & 0.10 & 0.39 & 10 \\
\hline\hline
\end{tabular}
\end{center}
\end{table}

\section{Chiral condensate with the UV cutoff}
\begin{figure}[t]
\begin{center}
\includegraphics[scale=1.5]{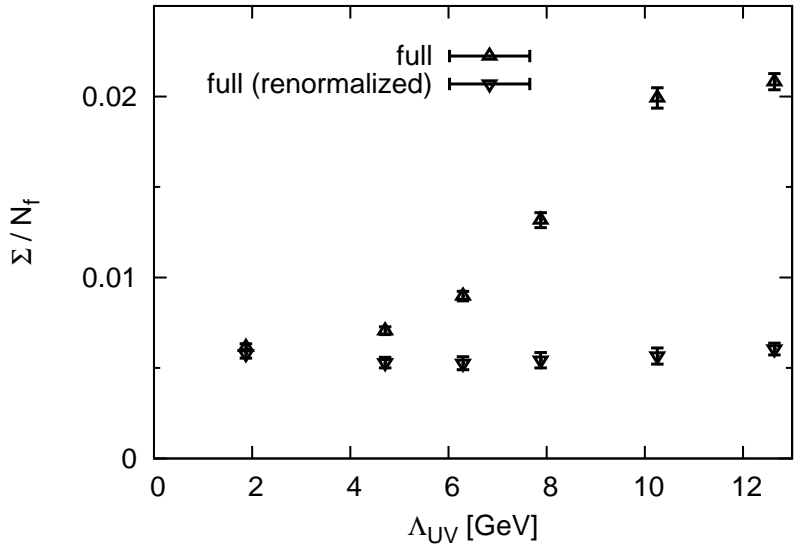}
\caption{\label{fig6-1}
The chiral condensate $\Sigma = -a^3\langle \bar{q}q \rangle$ with the UV cutoff $\Lambda_{\rm UV}$
The quark mass is $ma=0.01$.
The ``renormalized'' chiral condensate is multiplied by the renormalization factor $Z_S$.
}
\end{center}
\end{figure}

First, we show the chiral condensate with the UV cutoff in Fig.~\ref{fig6-1}.
Since there is no significant difference between the quenched and full QCD results, we plot only the full QCD result.
The right-side point at $\Lambda_{\rm UV}\simeq 12.5$ GeV is the result of original lattice QCD without the momentum cutoff.

Although spontaneous chiral symmetry breaking is expected to be caused by nonperturbative gluons, the chiral condensate is drastically changed by the UV cutoff.
However, as shown below, this is mainly because the chiral condensate is a renormalization-group variant and UV-diverging quantity.
It is dressed by perturbative gluons and its value strongly depends on the UV regularization.
In standard lattice QCD, the perturbative contribution is several orders of magnitude larger than the nonperturbative core of the condensate, as shown in Section 4.1.

To estimate the effect of renormalization, we calculate a renormalization factor, so-called a Z-factor, nonperturbatively \cite{Ma95,Ao99}.
The renormalization factor $Z_O(k)$ is determined from the amputated Green function of the quark bilinear operator $O$.
The renormalization condition is imposed as
\begin{eqnarray}
Z_O (k) Z^{-1}_q (k) \Gamma_O (k)=1,
\end{eqnarray}
where
\begin{eqnarray}
\Gamma_O (k) &\equiv& \frac{1}{16N_c} {\rm tr} [ S_q^{-1}(k) G_O (k) S_q^{-1}(k) P^\dagger_O]\\
G_O (k) &\equiv& \langle q(k) O \bar{q}(k) \rangle \\
S_q(k) &\equiv& \langle q(k) \bar{q}(k) \rangle.
\end{eqnarray}
$P_O$ is the appropriate projection operator.
The wave-function renormalization factor $Z_q^{1/2}(k)$ of the quark field is obtained from the conserved vector current, i.e., $Z_V(k)=1$.
Note that $k$ is the momentum of the quark field, not the momentum of the gluon field.

We calculate the renormalization factor $Z_S(k)$ of the scalar operator $O={\bar q}q$, and plot the renormalized chiral condensate $Z_S(5\ {\rm GeV}) \times \Sigma$ in Fig.~\ref{fig6-1}.
The renormalized chiral condensate is almost independent of the UV cutoff.
As the UV gluon is removed by the UV cutoff, the bare chiral condensate approaches the renormalized one.
This means that the drastic change by the UV cutoff is well explained in terms of renormalization.

\section{Chiral condensate with the IR cutoff}

\begin{figure}[p]
\begin{center}
\includegraphics[scale=1.5]{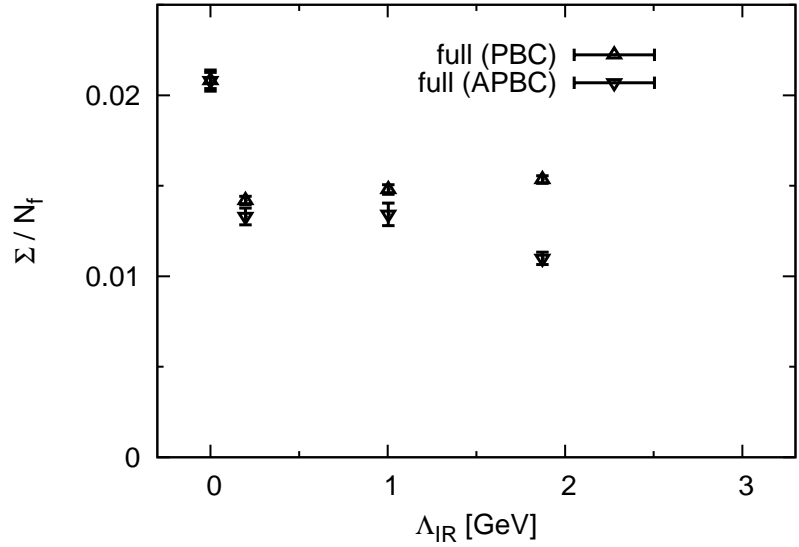}
\caption{\label{fig6-2}
The full QCD result of the chiral condensate $\Sigma = -a^3\langle \bar{q}q \rangle$ with the IR cutoff $\Lambda_{\rm IR}$.
The quark mass is $ma=0.01$.
PBC and APBC mean periodic and antiperiodic boundary conditions, respectively.
}
\end{center}
\end{figure}

\begin{figure}[p]
\begin{center}
\includegraphics[scale=1.5]{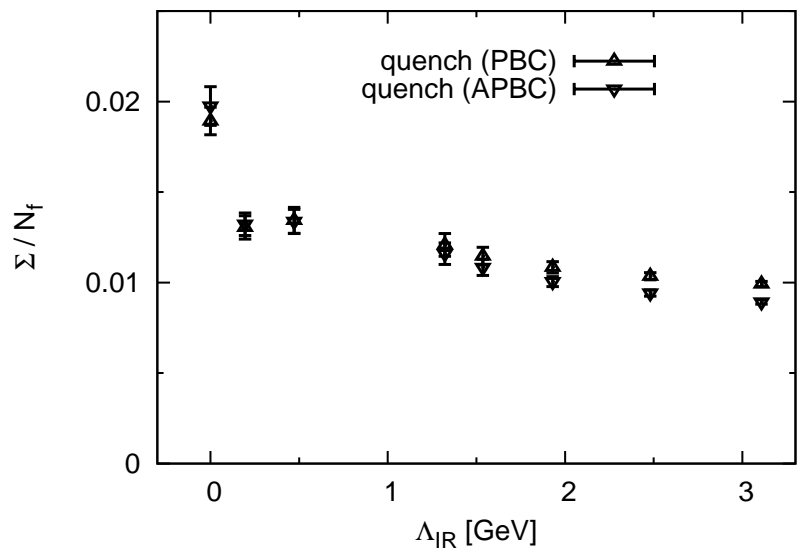}
\caption{\label{fig6-3}
The quenched QCD result of the chiral condensate with the IR cutoff.
}
\end{center}
\end{figure}

Second, we analyze the chiral condensate with the IR cutoff.
We show the full QCD result in Fig.~\ref{fig6-2} and the quenched QCD result in Fig.~\ref{fig6-3}.
The quark mass is $ma=0.01$ in both calculations.
In the case of the IR cutoff, the chiral condensate does not show the drastic change.
Thus, we expect the physical contribution to spontaneous chiral symmetry breaking instead of an artifact of renormalization.

When the IR gluon is removed, the effective quark mass would be reduced especially  at a large distance.
Thus, we must pay attention to the finite-volume effect in $\Lambda_{\rm IR}>0$, even though our lattice volume is large enough at $\Lambda_{\rm IR}=0$.
We estimate the finite-volume effect by changing boundary conditions of the quark propagator \cite{Do04}.
In Fig.~\ref{fig6-2} and Fig.~\ref{fig6-3}, ``PBC'' and ``APBC'' mean periodic and antiperiodic boundary conditions, respectively.
Since the result is independent of the boundary conditions if the lattice volume is large enough, the difference between these data should be understood as the finite-volume effect.
As seen from Fig.~\ref{fig6-2}, the $16^3 \times 32$ lattice of full QCD suffers from the finite-volume effect in $\Lambda_{\rm IR}>1.0$ GeV.
From Fig.~\ref{fig6-3}, the finite-volume effect is fairly small for the $32^4$ lattice of quenched QCD,
although it gradually grows in $\Lambda_{\rm IR}>1.5$ GeV.

Both in Fig.~\ref{fig6-2} and Fig.~\ref{fig6-3}, the chiral condensate suddenly gets small around $\Lambda_{\rm IR}=0$.
This jump around $\Lambda_{\rm IR}=0$ is caused by cutting only the zero-momentum link variable ${\tilde U}_{\mu}(0)$.
Despite the change at a single point $p^2=0$, the chiral condensate is about 40\% reduced.
Such a large change is not observed in removing other low-momentum components.
Therefore, the zero-momentum gluon is special and it possesses a major contribution to the chiral condensate.
Note that ``zero momentum'' on momentum-space lattice corresponds to the deep-IR region which is roughly $\sqrt{p^2}<a_p$ in the continuum.

The zero-momentum link variable corresponds to a spatially-uniform gauge background.
In general, the non-Abelian gauge field could have a nontrivial effect even in spatially-uniform case, unlike the Abelian gauge field. 
Our result actually suggests that the zero-momentum gauge field contributes to the chiral condensate. 
Note, however, that it is nontrivial whether spontaneous chiral symmetry breaking occurs only by the spatially-uniform gauge background. 

In large $\Lambda_{\rm IR}$, since the lattice volume of full QCD is not large enough, we analyze the quenched QCD result in Fig.~\ref{fig6-3}.
When the ``nonzero-momentum'' gluon of $\sqrt{p^2}\ge a_p$ is removed by the IR cutoff, the chiral condensate gradually decreases.
Thus, not only the ``zero-momentum'' gluon but also the ``nonzero-momentum'' gluon contributes to the chiral condensate.
The chiral condensate continues to decrease even in $\Lambda_{\rm IR}>1.5$ GeV.
Although it is difficult to perform an accurate analysis in large $\Lambda_{\rm IR}$ because of the finite-volume effect, we can see that the chiral condensate is affected by the gluon in the intermediate-momentum region.

\begin{figure}[t]
\begin{center}
\includegraphics[scale=1.5]{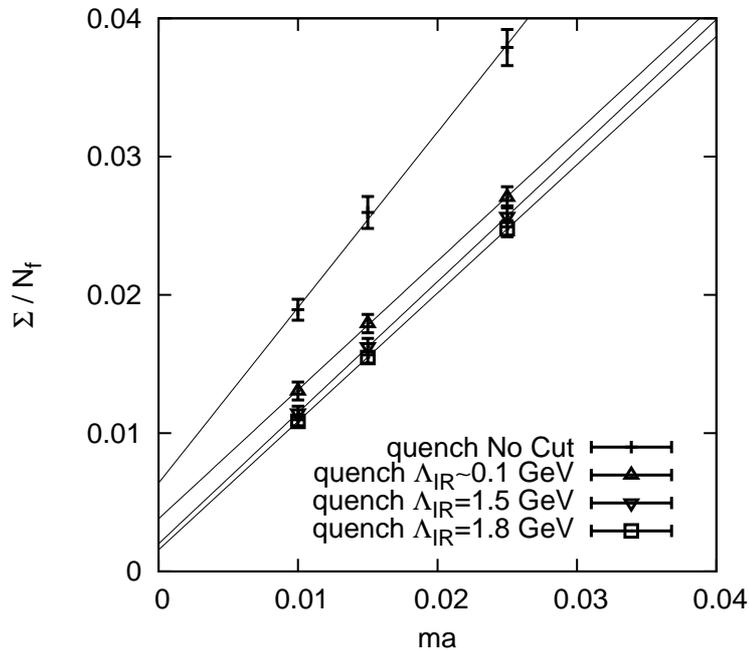}
\caption{\label{fig6-4}
The chiral extrapolation of the chiral condensate $\Sigma = -a^3\langle \bar{q}q \rangle$.
$\Lambda_{\rm IR}\sim 0.1$ GeV corresponds to the cutoff for the zero-momentum link variable.
}
\end{center}
\end{figure}

Next, we consider the chiral extrapolation of the chiral condensate.
When the bare quark mass $m$ is small, the chiral condensate is expanded as a function of $m$, as
\begin{eqnarray}
\Sigma (m) = \Sigma (0) + ma \Sigma'(0) + \cdots,
\end{eqnarray}
where $\Sigma'(m) \equiv \partial \Sigma (m)/ \partial ma$.
$\Sigma (0)$ represents spontaneous chiral symmetry breaking in the chiral limit.
We fit the quenched QCD result by the linear extrapolation function $\Sigma (0) + ma \Sigma'(0)$.
The fitting result is shown in Fig.~\ref{fig6-4} and Table \ref{tab6-2}.
Note that the data of ``$\Lambda_{\rm IR}\sim 0.1$ GeV'' corresponds to the smallest IR cutoff, which cuts only the zero-momentum link variable, and so the value ``0.1 GeV'' itself is not so meaningful.

As stated above, when the zero-momentum gluon field is removed, the chiral condensate is largely changed.
$\Sigma (0)$ is about 40\% reduced and $\Sigma'(0)$ is about 30\% reduced.
As for the nonzero-momentum gluon, the extrapolating line moves down parallel by the IR cutoff.
$\Sigma (0)$ is gradually reduced and $\Sigma'(0)$ is almost unchanged.
This indicates that the nonzero-momentum gluon has small but finite contribution to spontaneous chiral symmetry breaking.

In Fig.~\ref{fig6-4}, our result suggests another interesting possibility.
At least within the present numerical accuracy, the chiral condensate in the chiral limit remains finite at $\Lambda_{\rm IR}=1.5$ GeV, which is the relevant energy scale of color confinement.
If this is true, this means that the relevant energy scale of spontaneous chiral symmetry breaking is larger than that of color confinement at zero temperature.
Unfortunately, however, we cannot make a decisive statement because of systematic error of the chiral extrapolation.
For more conclusive answer, we need the full QCD calculation very close to the chiral limit, while the finite-volume effect is severely crucial in large $\Lambda_{\rm IR}$ and small $m$.

\begin{table}[t]
\begin{center}
\caption{\label{tab6-2}
The fitting result of the chiral extrapolation in Fig.~\ref{fig6-4}.
The extrapolation function is $\Sigma (0) + ma \Sigma'(0)$.
}
\begin{tabular}{ccc}
\hline\hline
$\Lambda_{\rm IR}$ [GeV] & $\Sigma (0)$ & $\Sigma'(0)$\\
\hline
0       & 0.00639(81) & 1.269(53)\\
$\sim 0.1$ & 0.00380(27) & 0.933(16)\\
1.5 & 0.00200(7)  & 0.948(4) \\
1.8 & 0.00155(2)  & 0.929(1) \\
\hline\hline
\end{tabular}
\end{center}
\end{table}

The Banks-Casher relation states that the chiral condensate is related to the spectral density $\rho (\lambda)$ of the Dirac operator as
\begin{eqnarray}
\Sigma = \pi \rho(0) ,
\end{eqnarray}
in the chiral limit \cite{Ba80}.
The spectral density of the Dirac operator is given in infinite volume as
\begin{eqnarray}
\rho (\lambda) = \lim _{V\to \infty} \frac{1}{V} \sum_k \delta (\lambda -\lambda_k),
\end{eqnarray}
and the eigenvalue of the Dirac operator is $i\lambda_k$.
The zero mode of the quark field is directly related to spontaneous chiral symmetry breaking from this relation.
In contrast, the gluon field is nontrivially related to spontaneous chiral symmetry breaking.
Our result presents the connection between the energy-momentum component of gluons and the zero mode of quarks. i.e., the energy scale of gluons and quarks.
We directly analyze the Dirac spectrum in the next chapter.

\chapter{Dirac spectrum}
\thispagestyle{headings}

In this chapter, we analyze the eigenvalue spectrum of the Dirac operator.
Although the Dirac equation is written in a simple form, the Dirac spectrum tells us nontrivial properties of QCD.
In particular, zero modes of the Dirac operator have special roles on nonperturbative properties induced by the gluon field, e.g., spontaneous chiral symmetry breaking and topological charge.
Moreover, the low-lying Dirac spectrum shows the universal behavior of disordered systems, which is described by chiral random matrix theory.
It is important to clarify how the gluon field induces these properties of the Dirac spectrum.

We computed the low-lying eigenvalues of the massless staggered Dirac operator in quenched lattice QCD.
The parameters of the gauge configurations are listed in Table \ref{tab7-1}.
We here adopted the 1-loop improved Symanzik gauge action with tadpole improvement \cite{Lu85}.
The tadpole coefficient is defined as the fourth-root of the average plaquette value \cite{Le93}.
We generated four kinds of gauge configurations.
The $12^4$, $16^4$, and $20^4$ lattices are used to analyze the eigenvalues and the chiralities, and the $8^4$ lattice is used to discuss the connection with chiral random matrix theory.

\begin{table}[h]
\renewcommand{\tabcolsep}{1pc} 
\renewcommand{\arraystretch}{1} 
\caption{\label{tab7-1}
Simulation parameters of the gauge configurations.
The notation is the same as in Table \ref{tab3-1}.
}
\begin{center}
\begin{tabular}{ccccccc}
\hline\hline
& $\beta$ & $V$ [$a^4$] & $a$ [fm] & $a_p$ [GeV] & $N_{\rm conf}$\\
\hline
Quenched (improved) & 9.0 & $20^4$ & 0.07 & 0.89 & 50 \\
Quenched (improved) & 8.6 & $16^4$ & 0.08 & 0.97 & 50 \\
Quenched (improved) & 8.3 & $12^4$ & 0.11 & 0.94 & 50 \\
Quenched (improved) & 7.9 & $8^4$  & 0.16 & 0.97 & 5000 \\
\hline\hline
\end{tabular}
\end{center}
\end{table}

\section{Eigenvalue and chirality}

The $j$-th eigenmode of the Dirac operator is given by
\begin{eqnarray}
D \psi_j = i \lambda_j \psi_j.
\end{eqnarray}
Because the massless staggered Dirac operator is anti-Hermitian, the eigenvalues always appear in pairs, $\pm i\lambda_j$.
We only have to calculate the positive eigenvalues $\lambda_j > 0$.
It should be understood that the negative eigenvalues with the same magnitude exist.

As explained in Section 2.1, the lattice staggered fermion breaks the flavor symmetry, i.e., the full SU($N_f$) chiral symmetry.
As a result, the unimproved staggered Dirac operator fails to reproduce nontrivial topological sectors \cite{Go99,Da99a,Da99b}.
This problem is approximately settled by improving the lattice gauge action and the lattice fermion action \cite{Du04,Fo04a,Wo05,Fo05}.
For the improved gauge action, we adopted the 1-loop improved Symanzik gauge action with tadpole improvement \cite{Lu85}.
For the improved staggered Dirac operator, we adopted the FAT7$\times$ASQ operator \cite{Fo04b}.
The FAT7$\times$ASQ operator is constructed from the FAT7 link variable and the ASQ operator \cite{Or99}.
The spatial boundary conditions are periodic, and the temporal boundary condition is antiperiodic.

\begin{figure}[t]
\begin{center}
\includegraphics[scale=1.2]{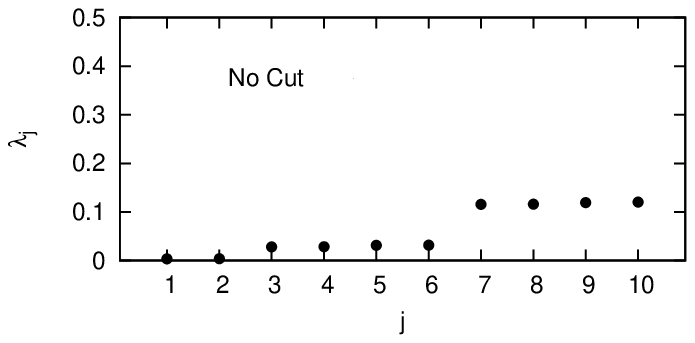}
\includegraphics[scale=1.2]{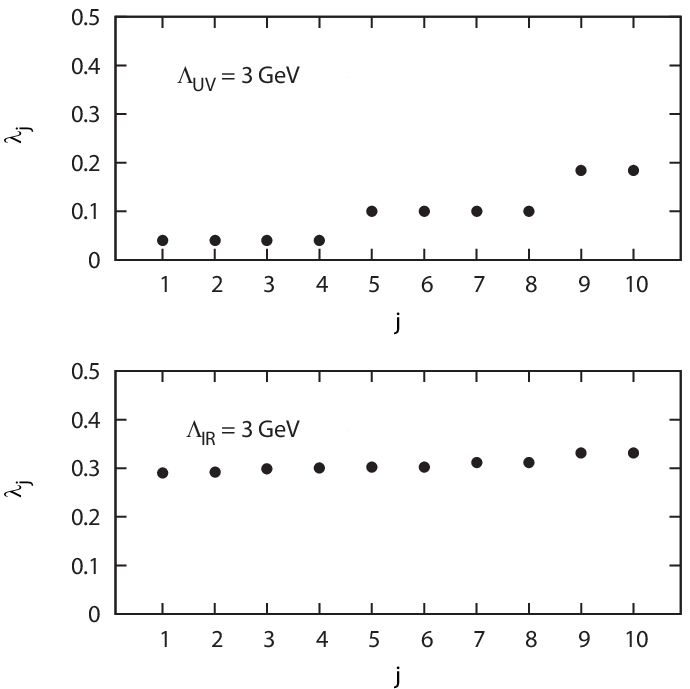}
\caption{\label{fig7-1}
The eigenvalue spectrum of the FAT7$\times$ASQ Dirac operator.
Only low-lying 10 positive eigenvalues of one typical gauge configuration are shown.
}
\end{center}
\end{figure}

In the top of Fig.~\ref{fig7-1}, we show the low-lying 10 positive eigenvalues of one typical gauge configuration of $\beta=8.6$.
This improved operator strongly suppresses the flavor-symmetry breaking lattice artifact.
The four-fold degeneracy, i.e., the four-flavor symmetry, is satisfied with high precision.
In addition, zero eigenvalues are approximately reproduced.
Such zero modes are called as ``would-be'' or ``near'' zero modes.
Because the negative eigenvalues with the same magnitudes exist, the topological charge of this gauge configuration is $Q=1$.

In Fig.~\ref{fig7-1}, we also show the Dirac spectrum of the same gauge configuration with $\Lambda_{\rm UV}=3$ GeV and $\Lambda_{\rm IR}=3$ GeV.
For $\Lambda_{\rm UV}=3$ GeV, while all the eigenvalues change, the four-fold degeneracy is still satisfied.
The would-be zero modes disappear.
This means that the topological charge is broken.
For $\Lambda_{\rm IR}=3$ GeV, the low-lying 10 eigenvalues become close to the eigenvalue of free fermions, which is about 0.39 for this case.
Also in this case, the topological charge becomes zero.

\begin{figure}[p]
\begin{center}
\includegraphics[scale=1.2]{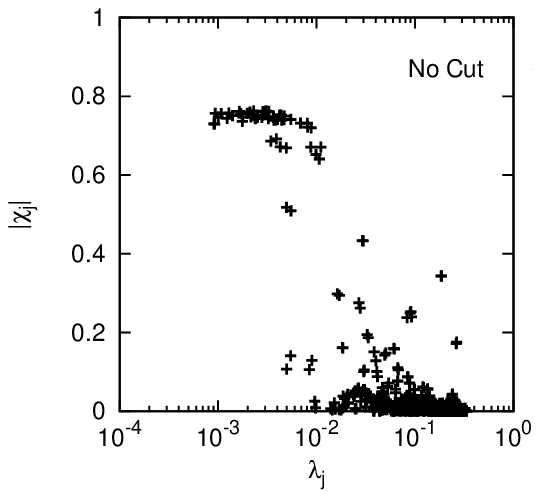}
\includegraphics[scale=0.9]{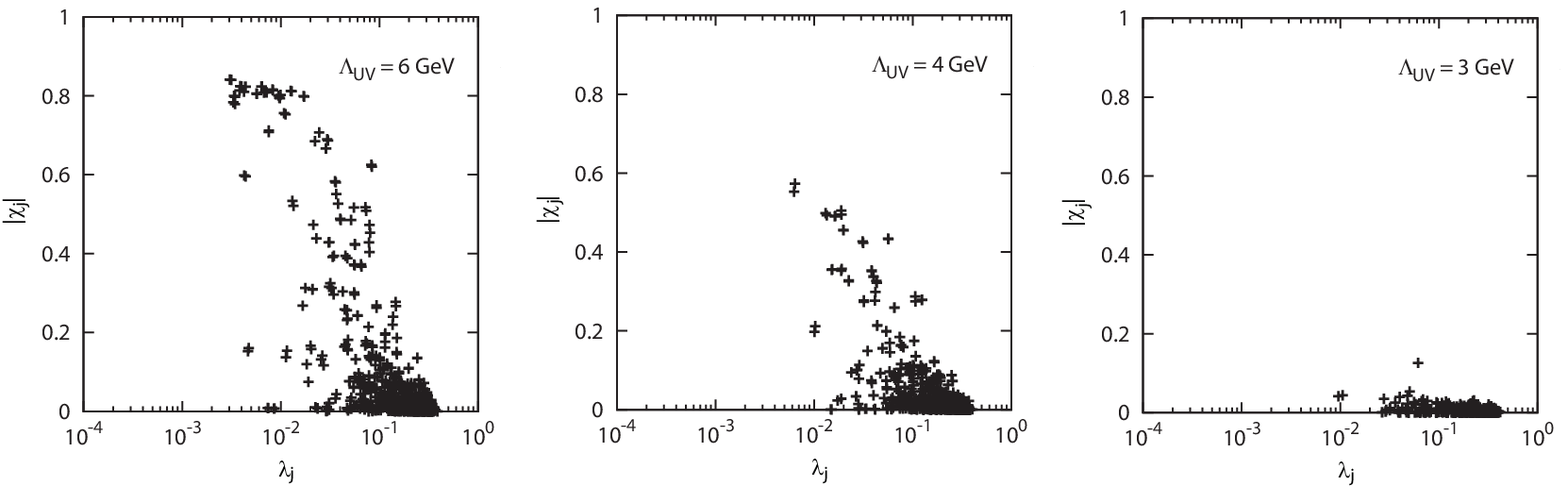}
\includegraphics[scale=0.9]{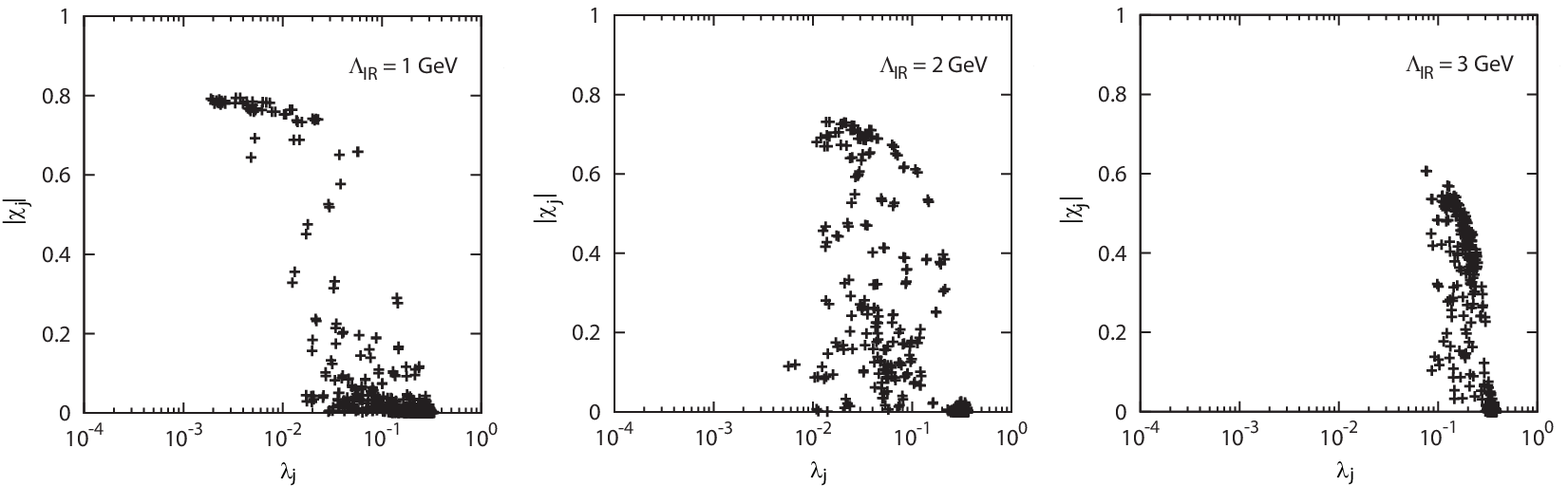}
\caption{\label{fig7-2}
The scattering plot of the eigenvalue $\lambda_j$ and the absolute value of the chirality $\chi_j$.
The low-lying 30 eigenvalues of 50 gauge configurations are plotted.
The calculation is done on $16^4$ lattice with $\beta=8.6$.
}
\end{center}
\end{figure}

Next, we discuss the correlation between the eigenvalue and the chirality.
The chirality is defined as
\begin{eqnarray}
\chi_j = \psi^\dagger_j \gamma_5 \psi_j.
\end{eqnarray}
For nonzero eigenvalues, their chiralities are zero.
On the other hand, for zero eigenvalues, their chiralities are $+1$ or $-1$.
In the top of Fig.~\ref{fig7-2}, we show the scattering plot of all the low-lying 30 eigenvalues of 50 gauge configurations of $\beta=8.6$.
There exists a clear separation between the nonzero modes, which is $|\chi_j|\simeq 0$, and the would-be zero modes, which is $|\chi_j|\simeq 0.8$.
The reason why the chirality deviates from $\pm 1$ is that it requires a renormalization \cite{Sm88}.

In the middle of Fig.~\ref{fig7-2}, we show the scattering plot with the UV cutoff $\Lambda_{\rm UV} =6$, 4, and 3 GeV.
As the high-momentum component is removed, the eigenvalue spectrum is gradually changed.
For $\Lambda_{\rm UV}=3$ GeV, all eigenmodes are $|\chi_j|\simeq 0$.
Therefore, all gauge configurations fall into the trivial topological sector.
In the bottom of Fig.~\ref{fig7-2}, we show the scattering plot with the IR cutoff $\Lambda_{\rm IR} =1$, 2, and 3 GeV.
When the low-momentum component is removed, many of the low-lying eigenvalues approach the free eigenvalue, which is $\lambda_j \simeq 0.39$ and $|\chi_j|\simeq 0$.
Compared to the case of the UV cutoff, the changes of the eigenvalues are sensitive to the IR cutoff.
This result is consistent with our expectation that the low-lying Dirac eigenmode interacts with the low-momentum gluon.
For $\Lambda_{\rm IR}=3$ GeV, we cannot identify would-be zero modes since all eigenvalues are almost degenerate.
From these results, except for the very-high-momentum components, the relatively broad region of momentum components of the gluon field is related to the low-lying Dirac spectrum.

When the momentum cutoff is imposed, the clear separation between would-be zero modes and nonzero modes is lost.
This is because the staggered Dirac operator does not have exact zero modes.
This situation will be improved by the overlap operator, which has exact zero modes \cite{Ne98,Ad10}.

\section{Topological charge}
The numbers of zero modes determine the topological charge $Q$ of the background gauge field through the Atiyah-Singer index theorem \cite{At68}.
The index theorem states
\begin{eqnarray}
n_R - n_L = N_f Q,
\end{eqnarray}
where $n_R$ and $n_L$ are the numbers of right-handed and left-handed zero modes, respectively \cite{At68}.

From the number of would-be zero modes, we extracted the topological charge $Q$.
We set the criterion for would-be zero modes as $|\chi_j|> 0.6$, and for nonzero modes as $|\chi_j|< 0.6$.
This criterion clearly separates would-be zero modes in the original case, as shown in the top of Fig.~\ref{fig7-2}.
For the cases in which the clear separation between would-be zero modes and nonzero modes is lost, this criterion could be subtle.
The results are summarized in Table \ref{tab7-2}.

For a consistency check, we also extracted the topological charge in another way.
We directly calculated the topological charge by
\begin{eqnarray}
Q=\frac{1}{32\pi^2}\int d^4x \sum_{\mu\nu\sigma\rho} \epsilon_{\mu\nu\sigma\rho} {\rm Tr}[ F_{\mu\nu}(x) F_{\sigma\rho}(x) ].
\end{eqnarray}
Its discretized form is
\begin{eqnarray}
Q=-\frac{1}{512\pi^2}\sum_x \sum_{\mu\nu\sigma\rho=\pm 1}^{\pm 4} \epsilon_{\mu\nu\sigma\rho} {\rm Tr}[ U_{\mu\nu}(x) U_{\sigma\rho}(x) ],
\label{eqQG}
\end{eqnarray}
where $U_{\mu\nu}(x)$ is the plaquette.
We measured this topological charge after the APE smearing steps \cite{Al87,Da99b}.
The smearing coefficient and the number of the smearing step are set to 7.0 and 200, respectively.
The results are shown in the parentheses in Table \ref{tab7-2}.
The obtained topological charge is roughly consistent with the estimate by the would-be zero modes, especially in the original lattice QCD, as expected.
These two estimates deviate from each other in some cases with the momentum cutoff, e.g., $\Lambda_{\rm UV}=4$ GeV.
This is partly because it is difficult to identify the would-be zero modes in these cases, and partly because the APE smearing steps change the local topological structure.

We can see that the topological charge is destroyed by removing the broad momentum region of the gauge field.
This behavior is seen in the both cases of the UV and IR cutoffs.
Thus, there is no typical energy scale of the gauge field inducing the topological charge.
This is different from the case of color confinement, which is induced only by the narrow low-momentum component of the gauge field, as shown in Chapter 3.

In Table \ref{tab7-3}, we show the numerical results calculated in the Coulomb gauge, instead of the Landau gauge.
The results are almost consistent with those of the Landau gauge.

\begin{table}[h]
\begin{center}
\renewcommand{\tabcolsep}{0.5pc} 
\renewcommand{\arraystretch}{1} 
\caption{\label{tab7-2}
Topological charge $Q$ estimated by would-be zero modes.
The estimate by Eq.~(\ref{eqQG}) is also shown in the parenthesis.
The results of the 50 gauge configurations with $\beta=8.6$ are shown.
}
\begin{tabular}{cccccc}
\hline\hline
$V$ [$a^4$] & Data & $|Q|=0$ & $|Q|=1$ & $|Q|=2$ & $|Q|=3$\\
\hline
$16^4$ &No Cut & 28 (29) & 18 (17) & 4 (4) & 0 (0) \\
&$\Lambda_{\rm UV}=6$ GeV & 35 (27) & 12 (17) & 3 (6) & 0 (0) \\
&$\Lambda_{\rm UV}=4$ GeV & 50 (44) &  0 (6)  & 0 (0) & 0 (0) \\
&$\Lambda_{\rm UV}=3$ GeV & 50 (50) &  0 (0)  & 0 (0) & 0 (0) \\
&$\Lambda_{\rm IR}=1$ GeV & 29 (27) & 17 (19) & 4 (4) & 0 (0) \\
&$\Lambda_{\rm IR}=2$ GeV & 32 (43) & 13 (7)  & 5 (0) & 0 (0) \\
&$\Lambda_{\rm IR}=3$ GeV & 49 (50) &  1 (0)  & 0 (0) & 0 (0) \\
\hline
$20^4$ &No Cut & 28 (28) & 22 (22) & 0 (0) & 0 (0) \\
&$\Lambda_{\rm UV}=4$ GeV & 50 (50) &  0 (0)  & 0 (0) & 0 (0) \\
&$\Lambda_{\rm UV}=3$ GeV & 50 (50) &  0 (0)  & 0 (0) & 0 (0) \\
\hline
$12^4$ &No Cut & 28 (28) & 20 (20) & 2 (2) & 0 (0) \\
&$\Lambda_{\rm UV}=4$ GeV & 49 (44) &  1 (6)  & 0 (0) & 0 (0) \\
&$\Lambda_{\rm UV}=3$ GeV & 50 (50) &  0 (0)  & 0 (0) & 0 (0) \\
\hline\hline
\end{tabular}

\caption{\label{tab7-3}
Topological charge $Q$ in the Coulomb gauge.
}
\begin{tabular}{cccccc}
\hline\hline
$V$ [$a^4$] & Data & $|Q|=0$ & $|Q|=1$ & $|Q|=2$ & $|Q|=3$\\
\hline
$16^4$ &No Cut & 28 (29) & 18 (17) & 4 (4) & 0 (0) \\
&$\Lambda_{\rm IR}=1$ GeV & 29 (27) & 17 (20) & 4 (3) & 0 (0) \\
&$\Lambda_{\rm IR}=2$ GeV & 35 (31) & 13 (14) & 1 (5) & 1 (0) \\
&$\Lambda_{\rm IR}=3$ GeV & 49 (50) &  1 (0)  & 0 (0) & 0 (0) \\
\hline\hline
\end{tabular}
\end{center}
\end{table}

\section{Connection with chiral random matrix theory}
Another interesting property of the low-lying Dirac spectrum is that it is successfully described by chiral random matrix theory.
Chiral random matrix theory is an effective theory which describes the universal behavior of disordered systems \cite{Sh93,Gu98,Ve00}.
Chiral random matrix theory suggests that the low-lying Dirac spectrum depends only on the global symmetries of the system, and not on the details of the microscopic structure.
Once the validity of chiral random matrix theory is guaranteed, chiral random matrix theory enables us to predict the low-lying eigenvalue distribution by a simple and universal function.

We compared the lattice data with a prediction of chiral random matrix theory, using the gauge configurations of $\beta=7.9$.
We analyzed the $Q=0$ sector with the criterion for nonzero modes as $|\chi_j|< 0.1$.

\begin{figure*}[p]
\begin{center}
\includegraphics[scale=1.8]{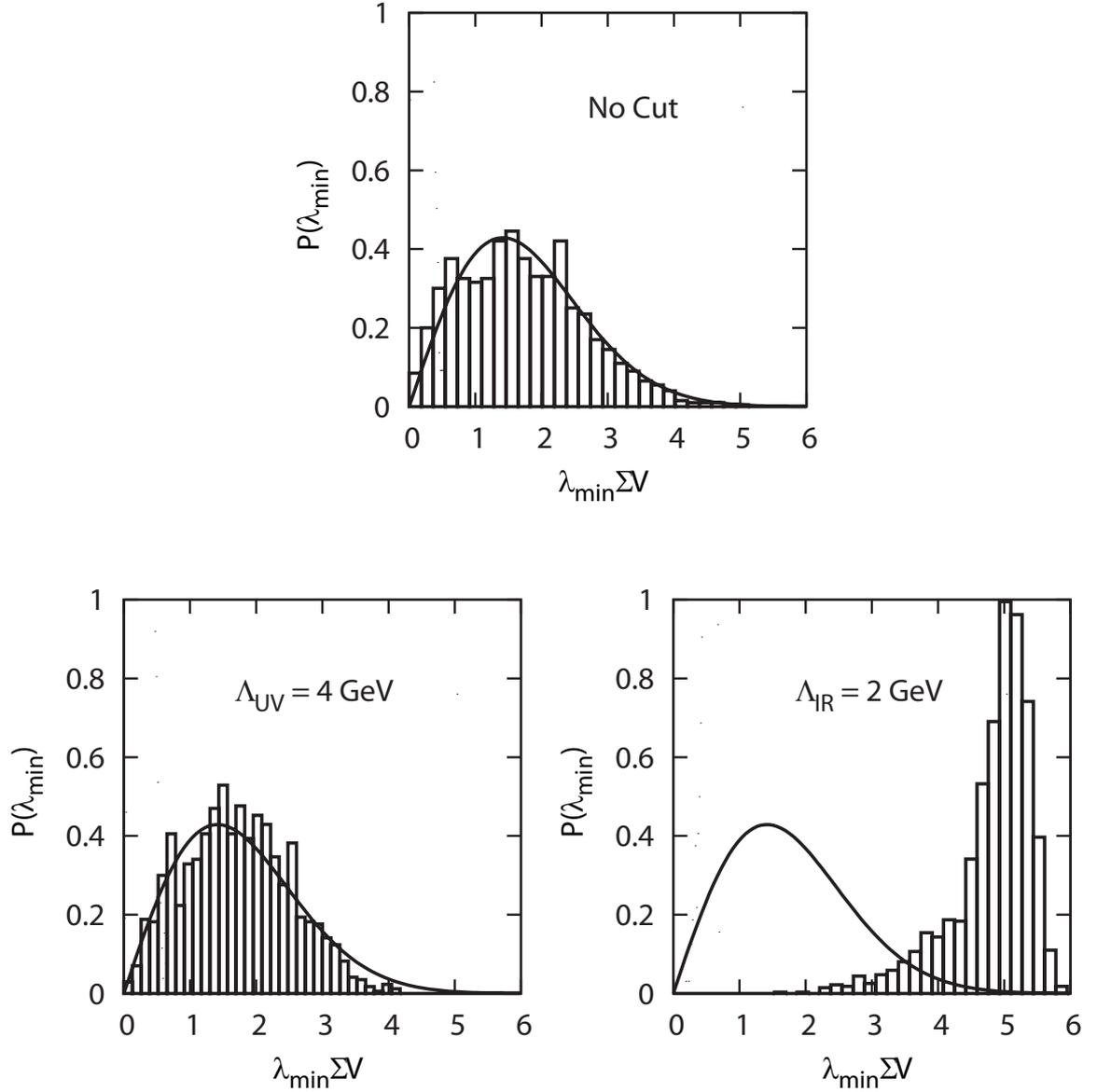}
\caption{\label{fig7-3}
The probability distribution of the lowest eigenvalue $\lambda_{\rm min}$ in the $Q=0$ sector.
The histograms are the lattice data of $8^4$ lattice with $\beta=7.9$.
The solid curves for No Cut and $\Lambda_{\rm UV}=4$ GeV are the best-fit functions of Eq.~(\ref{eqP}) from chiral random matrix theory.
}
\end{center}
\end{figure*}

We calculated the probability distribution of the lowest nonzero eigenvalue $\lambda_{\rm min}$.
In chiral random matrix theory, the probability distribution of the lowest eigenvalue is given as
\begin{eqnarray}
P(\lambda_{\rm min}) &=& \frac{z_{\rm min}}{2} e^{-z_{\rm min}^2/4}
\label{eqP}\\
z_{\rm min} &\equiv& \lambda_{\rm min} \Sigma V
\end{eqnarray}
for the SU(3) quenched gauge field with $Q=0$ \cite{Fo93}.
If one directly calculates the chiral condensate in infinite volume, this function is parameter free.
In this study, we treated $\Sigma$ as a fitting parameter.
The lattice data and the best-fit result of Eq.~(\ref{eqP}) are shown in Fig.~\ref{fig7-3}.
In the original lattice QCD, the lattice data is well reproduced by chiral random matrix theory.
The best-fit value is $\Sigma \simeq 0.0022\times a^{-3}$.

For $\Lambda_{\rm UV}=4$ GeV, the qualitative behavior does not change.
The lattice data is still reproduced by Eq.~(\ref{eqP}).
Thus, the high-momentum gluon is irrelevant for the universality of the lowest eigenvalue distribution.
The best-fit value of $\Sigma$ is changed by the UV cutoff.
This is because the chiral condensate is renormalization-group variant and UV divergent.
By definition, the UV divergent quantity depends on the UV cutoff, although this might be irrelevant for phenomenology.

For $\Lambda_{\rm IR}=2$ GeV, the lowest eigenvalue distribution becomes to concentrate on the vicinity of the free eigenvalue.
The lattice data drastically deviates from Eq.~(\ref{eqP}).
Therefore, the low-momentum gluon is crucial for the validity of chiral random matrix theory.
In other words, the low-momentum component induces the strong-interacting and disordered nature of the gauge field.

In the calculation of the chiral condensate in Chapter 6, the zero-momentum component has a large contribution to the chiral condensate.
To check it, we calculated the Dirac spectrum with the small IR cutoff $\Lambda_{\rm IR}\sim 0.1$ GeV.
For $\Lambda_{\rm IR}\sim 0.1$ GeV, the lattice data can be fitted by Eq.~(\ref{eqP}) with the best-fit value $\Sigma \simeq 0.0019\times a^{-3}$.
The contribution of the zero-momentum component is indeed large, about 15\% of the total.
This is qualitatively consistent with the result in Chapter 6.
Note, however, that this is quantitatively different because the lattice action and the lattice spacing are different.

\chapter{Summary}
\thispagestyle{headings}

We have introduced the lattice framework to analyze the relation between a QCD phenomenon and the energy-momentum component of the gluon field.
In this framework, using the Fourier transformation of the link variable, we construct the momentum-space link variable, remove some energy-momentum components, and reconstruct the coordinate-space link variable.
This framework is applicable to many lattice QCD calculations.
We have applied this framework to the analysis of color confinement and spontaneous chiral symmetry breaking.

As for color confinement, we have analyzed the quark-antiquark potential, the color flux tube, and meson masses.
From the quantitative analysis of the quark-antiquark potential, we have found that the confinement potential is induced by the low-momentum component below 1.5 GeV in the Landau gauge.
The low-momentum component below 1.5 GeV produces the linear interquark potential, and the high-momentum component above 1.5 GeV produces the Coulomb-like interquark potential without confinement.
As one application of this result, by restricting the gluon field to the high-momentum component above 1.5 GeV, we can extract the color flux tube from the action density distribution.
In addition, when this low-momentum component is removed, the two quarks in mesons are not confined and become quasi-free.

As for spontaneous chiral symmetry breaking, we have analyzed the chiral condensate and the Dirac spectrum.
The chiral condensate is induced by the broad low-momentum region, even above 1.5 GeV.
This behavior is also observed in the analysis of the Dirac spectrum.
This dependence on the momentum cutoff is different from that of color confinement.
The present result suggests that spontaneous chiral symmetry breaking is induced by higher energy-momentum component of gluons, compared to color confinement.
This behavior is consistent with phenomenological models \cite{Ca79,Ma84}.
For more conclusive statement, we have to take into account many systematics, especially, the sea quark effects in the case of realistic QCD.

We have also analyzed topological charge in the two ways: the fermionic way through the Dirac zero mode and the field theoretical way through the gauge field distribution.
The momentum cutoff gradually affects the Dirac spectrum and the zero modes, and then changes the topological charge.
This is similar to the behavior of the chiral condensate.

In this framework, we need gauge fixing, and the obtained result depends on the gauge choice.
Nevertheless, this analysis is important in the following theoretical viewpoints.

First, the relevant energy scale, i.e., the energy-momentum component inducing a QCD phenomenon, would be useful for developing effective theories.
An effective theory has some cutoff to simplify physics or to reduce degrees of freedom.
Although the value of the cutoff should be based on some physical reasoning, its microscopic derivation is difficult in many cases.
The relevant energy scale obtained by lattice QCD would provides a physical reasoning for the cutoff value.
For example, by setting the UV cutoff to be 1.5 GeV in momentum integral, we can safely pick up the contribution to color confinement.
Further, the relevant energy scale determines degrees of freedom which appear in the effective field theory \cite{Br05}.

Second, this analysis reveals the relation between different QCD phenomena in terms of the energy-momentum component of the gauge field.
We studied the connection between color confinement and spontaneous chiral symmetry breaking.
Such connection would be important for understanding their underlying mechanisms.
Also, based on the connection between the confinement potential and the color flux tube, we succeeded in extracting the color flux tube from the gluon distribution.

Third, this framework has a technical advantage in numerical simulation.
As shown in Chapter 4, the statistical fluctuation of the action density is strongly suppressed by removing the UV components.
This is because the statistical fluctuation is mainly given by the high-frequency mode, i.e., the UV component of the link variable.
The statistical error is one major difficulty in lattice QCD.
This framework can reduce the statistical error without changing the target IR physics \cite{Ta09}. 

\chapter*{Acknowledgments}
\thispagestyle{headings}
The author is very grateful to the adviser, Hideo Suganuma, for all the discussions and collaborations in this work.
He kindly supported the research activity in graduate school.

The author would like to thank all the collaborators in other works, Teiji Kunihiro, Akira Ohnishi, Berndt M\"uller, Andreas Sch\"afer, Toru T. Takahashi, and Hideaki Iida.
The author also thanks the members of Nuclear Theory Groups in Kyoto University and in Yukawa Institute for Theoretical Physics.

The author is supported by Japan Society for the Promotion of Science and a Grant-in-Aid for Scientific Research [(C) No.~20$\cdot$363].

The lattice QCD simulations were carried out on Altix3700 BX2 and SX8 at YITP in Kyoto University, and SX8 at Osaka University.

\end{document}